\documentclass[iop,apj]{emulateapj}
\usepackage{graphicx,url}

\shorttitle{Li-Rich Giants}
\shortauthors{Kirby et al.}
\slugcomment{Accepted to ApJ on 2016 January 14}
\begin{document}

\newcommand{\ndup}{437}
\newcommand{\nulrgb}{1916}
\newcommand{\nulagb}{157}
\newcommand{\nmeasrgb}{141}
\newcommand{\nmeasagb}{13}
\newcommand{\nrichrgb}{4}
\newcommand{\nrichagb}{2}
\newcommand{\nmemrgb}{2057}
\newcommand{\nmemagb}{170}
\newcommand{\nmeas}{154}
\newcommand{\nul}{2073}
\newcommand{\nulsig}{1673}
\newcommand{\nsigrgb}{1700}
\newcommand{\nsigagb}{127}
\newcommand{\nsig}{1827}
\newcommand{\nnormalrgb}{1696}
\newcommand{\nnormalagb}{125}
\newcommand{\nnormal}{1821}
\newcommand{\nmem}{2227}
\newcommand{\nrich}{6}
\newcommand{\fracrgb}{0.2}
\newcommand{\fracagb}{1.6}
\newcommand{\fractot}{0.3}
\newcommand{\fracrgberr}{0.1}
\newcommand{\fracagberr}{1.1}
\newcommand{\fractoterr}{0.1}
\newcommand{\proboverlap}{6.6}
\newcommand{\ncluster}{25}
\newcommand{\nmask}{75}
\newcommand{\totexptime}{72}
\newcommand{\lii}{Li$\,${\sc i}}
\newcommand{\liin}{Li$\,${\sc i}$\,\lambda$6707}
\newcommand{\baii}{Ba$\,${\sc ii}}
\newcommand{\baiin}{Ba$\,${\sc ii}$\,\lambda$6496}

\title{Lithium-Rich Giants in Globular Clusters\altaffilmark{*}}

\author{Evan~N.~Kirby\altaffilmark{1},
  Puragra~Guhathakurta\altaffilmark{2},
  Andrew~J.~Zhang\altaffilmark{3,4},
  Jerry~Hong\altaffilmark{5},
  Michelle~Guo\altaffilmark{4,6},
  Rachel~Guo\altaffilmark{6},
  Judith~G.~Cohen\altaffilmark{1},
  Katia~Cunha\altaffilmark{7,8}}

\altaffiltext{*}{The data presented herein were obtained at the
  W.~M.~Keck Observatory, which is operated as a scientific
  partnership among the California Institute of Technology, the
  University of California and the National Aeronautics and Space
  Administration. The Observatory was made possible by the generous
  financial support of the W.~M.~Keck Foundation.}
\altaffiltext{1}{California Institute of Technology, 1200 E.\ California Blvd., MC 249-17, Pasadena, CA 91125, USA}
\altaffiltext{2}{UCO/Lick Observatory and Department of Astronomy and Astrophysics, University of California, 1156 High St., Santa Cruz, CA 95064, USA}
\altaffiltext{3}{The Harker School, 500 Saratoga Ave., San Jose, CA 95129, USA}
\altaffiltext{4}{Stanford University, 450 Serra Mall, Stanford, CA 94305, USA}
\altaffiltext{5}{Palo Alto High School, 50 Embarcadero Rd., Palo Alto, CA, 94301, USA}
\altaffiltext{6}{Irvington High School, 41800 Blacow Rd., Fremont, CA 94538, USA}
\altaffiltext{7}{Observat{\' o}rio Nacional, S{\~ a}o Crist{\' o}v{\~ a}o Rio de Janeiro, Brazil}
\altaffiltext{8}{University of Arizona, Tucson, AZ 85719, USA}



\begin{abstract}

  
Although red giants deplete lithium on their surfaces, some giants are
Li-rich.  Intermediate-mass asymptotic giant branch (AGB) stars can
generate Li through the Cameron--Fowler conveyor, but the existence of
Li-rich, low-mass red giant branch (RGB) stars is puzzling.  Globular
clusters are the best sites to examine this phenomenon because it is
straightforward to determine membership in the cluster and to identify
the evolutionary state of each star.  In \totexptime\ hours of
Keck/DEIMOS exposures in \ncluster\ clusters, we found four Li-rich
RGB and two Li-rich AGB stars.  There were \nnormalrgb\ RGB and
\nnormalagb\ AGB stars with measurements or upper limits consistent
with normal abundances of Li.  Hence, the frequency of Li-richness in
globular clusters is $(\fracrgb \pm \fracrgberr)\%$ for the RGB,
$(\fracagb \pm \fracagberr)\%$ for the AGB, and $(\fractot \pm
\fractoterr)\%$ for all giants.  Because the Li-rich RGB stars are on
the lower RGB, Li self-generation mechanisms proposed to occur at the
luminosity function bump or He core flash cannot explain these four
lower RGB stars.  We propose the following origin for Li enrichment:
(1) All luminous giants experience a brief phase of Li enrichment at
the He core flash.  (2) All post-RGB stars with binary companions on
the lower RGB will engage in mass transfer.  This scenario predicts
that 0.1\% of lower RGB stars will appear Li-rich due to mass transfer
from a recently Li-enhanced companion.  This frequency is at the lower
end of our confidence interval.

\end{abstract}


\section{Introduction}
\label{sec:intro}

Lithium was created in the Big Bang at a concentration of about 0.5
parts per billion \citep{coc12}.  Since then, many of the Universe's
Li nuclei have been destroyed in nuclear burning because Li is
susceptible to proton capture at relatively low temperatures ($T
\gtrsim 2.5 \times 10^6$~K)\@.  Li burning occurs in the centers of
stars, but their surfaces are cool enough to preserve Li.  Therefore,
Li is observable only in stars with outer envelopes that have never
been fully mixed down to high temperatures.

\addtocounter{footnote}{-1} The atmospheres of most old, metal-poor
stars on the main sequence display the same amount of Li
\citep{spi82}.  This value, $A({\rm Li}) \sim 2.2$, is called the
Spite plateau.\footnote{$A({\rm Li}) = 12 + \log n({\rm Li})/n({\rm
    H})$, where $n({\rm Li})$ is the number density of Li atoms and
  $n({\rm H})$ is the number density of H atoms.}  However, the
plateau is significantly below the primordial value, $A({\rm Li}) =
2.72$ \citep{coc12}.  Although the factor of 2--4 drop in Li abundance
from the primordial value to the Spite plateau has been attributed to
atomic diffusion and turbulent transport below the convection zone on
the main sequence \citep{ric05,mel10} and convective overshoot on the
pre-main sequence \citep{fu15}, models of rotationally induced mixing
\citep[e.g.,][]{pin89} offer an explanation with less fine tuning.
\citet{pin92,pin99,pin02} showed that calibrating mixing parameters to
the Sun also explains Li depletion in other stars, including the mean
and dispersion of the Li abundance on the Spite plateau.  In addition,
the rotation models also explain the behavior of other light elements,
like Be and B \citep{del97,del98,boe05}.

In metal-rich stars, mixing more efficiently depletes surface lithium
than in metal-poor stars \citep[e.g.,][]{mel14,tuc15}.  Furthermore,
novae can generate Li for metal-rich, Population~I stars
\citep{rom99,taj15,izz15}.  As a result, the Spite plateau breaks down
at ${\rm [Fe/H]} \gtrsim -1.2$ \citep[e.g.,][]{che01}.  The constancy
of Li on the Spite plateau makes Li anomalies in metal-poor stars
readily apparent.  For example, some carbon-rich stars show
deficiencies in Li that can be explained by mass transfer from a
binary, Li-depleted companion \citep{mas12}.

However, it is more difficult to explain stars that are anomalous for
being {\it enhanced} in Li.  This is especially true for giant stars.
Stars at the main sequence turn-off experience a rapid drop in Li
abundance \citep{pil93,rya95,lin09b}.  As a low-mass star evolves on
to the red giant branch (RGB), its surface convection zone deepens
enough to dredge up material that has been processed through nuclear
fusion, including Li burning.  Although those regions are no longer
hot enough to burn Li, they were once hot enough to do so.  Hence, the
dredge-up brings up Li-depleted material while Li on the surface is
subducted into the star.  The dredge-up dilutes the surface Li
abundance to 5--10\% of its original value.  Models of dilution caused
by the dredge-up \citep{del90} explain the surface abundance of Li as
a function of the subgiant's increasing luminosity or decreasing
temperature.  When the star reaches a luminosity of $M_V \sim 0$, the
hydrogen-burning shell expands beyond the molecular weight boundary
established by the first dredge-up \citep{ibe68}.  ``Extra''
mixing---possibly thermohaline mixing
\citep{cha07,cha10,den10,wac11,ang12,lat15}---beyond the canonical
stellar model changes the surface composition for stars at the RGB
bump.  This mixing rapidly destroys any Li remaining in the red
giant's atmosphere.

Nonetheless, some giants are Li-rich \citep[see][]{wal69}.
\citet{cam55} and \citet{cam71} suggested a mechanism (the
``Cameron--Fowler conveyor'') for producing excess Li in the
atmospheres of giant stars.  The central nuclear processes for the
conveyor comprise the $pp$-II chain of hydrogen burning.

\begin{eqnarray}
p + p \: &\rightarrow& \: d + e^+ + \nu_e \\
d + p \: &\rightarrow& \: {}^{3}{\rm He} + \gamma \\
{}^{3}{\rm He} + {}^{4}{\rm He} \: &\rightarrow& \: {}^{7}{\rm Be} + \gamma \label{eq:ppii} \\
{}^{7}{\rm Be} + e^- \: &\rightarrow& \: {}^{7}{\rm Li} + \nu_e \label{eq:be} \\
{}^{7}{\rm Li} + p \: &\rightarrow& \: 2{}^{4}{\rm He} \label{eq:li}
\end{eqnarray}

\noindent
Reaction~\ref{eq:ppii} is very active (compared to the $pp$-I chain)
at temperatures around $2 \times 10^7$~K\@.  Li destruction,
reaction~\ref{eq:li}, is very efficient at $T \gtrsim 2.5 \times
10^6$~K\@.  Hence, ${}^7{\rm Li}$ will be destroyed as soon as it is
created in reaction~\ref{eq:be} unless ${}^7{\rm Be}$ can be brought
to cooler temperatures before it captures an electron.  Although the
half-life for reaction~\ref{eq:be} is 53 days under terrestrial
conditions, \citet{cam55} theorized that the scarcity of bound K-shell
electrons available for reaction~\ref{eq:be} at $T > 10^6$~K, where
${}^7{\rm Be}$ is almost entirely ionized, extends the half-life to
50--100 years.

The mixing that accompanies thermal pulses in intermediate-mass stars
on the second-ascent asymptotic giant branch (AGB) is deep enough to
reach the $pp$-II burning zone.  As a result, the Cameron--Fowler
conveyor is a plausible explanation for Li-rich AGB stars in the mass
range 4--7~$M_{\sun}$ \citep{sac92}.  In fact, Li-rich AGB stars are
not uncommon \citep{ple93,smi95}.  However, the convective envelopes
of first-ascent RGB stars and less massive AGB stars are not deep
enough to activate the conveyor.  Any excess Li in RGB stars must be a
result of processes outside of ``standard'' stellar evolution of
single stars with ordinary rotation rates.  \citet{sac99} called this
non-standard phenomenon extra deep mixing combined with ``cool bottom
processing.''  The mechanism for the mixing remains elusive.

Nonetheless, Li-rich red giants do exist.  \citet{kra99} discovered a
luminous red giant with $A({\rm Li}) = 3.0$ in the globular cluster
(GC) M3.  The star is unremarkable except for having over 1000 times
more Li than it should have, based on its position on the RGB\@.
Other GCs with Li-rich giants include M5 \citep[in a post-AGB
  Cepheid,][]{car98}, NGC~362 \citep{smi99,dor15a}, and M68
\citep{ruc11}.  \citet{kum09} and \citet{kum11} found over a dozen
Li-rich field K giants around solar metallicity.  They also found
tentative evidence for clustering of Li-rich giants around the red
clump, or horizontal branch (HB), where stars burn helium in their
cores after the He core flash at the tip of the RGB\@.  The idea that
the He core flash could activate the Cameron--Fowler conveyor was
bolstered by \citeauthor{sil14}'s (\citeyear{sil14}) discovery of a
Li-rich HB star whose He core burning was confirmed by
asteroseismological measurements from the Kepler spacecraft
\citep{gil10}.  \citet{mon14} also discovered a Li-rich, HB star in
the open cluster Trumpler~5, and \citet{ant13} found a Li-rich giant
in the open cluster NGC~6819 that is too faint to be on the HB or
AGB\@.  This star is particularly interesting for showing
asteroseismic anomalies that could indicate rotationally induced
mixing, which in turn could generate Li \citep[e.g.,][]{den12}.
Indeed, \citet{car15} found that the star is rotating rapidly for a
red giant, but puzzlingly, they did not find any additional evidence
for deep mixing.

Metal-rich stars can have a complicated evolution of Li, as
illustrated by the $\sim 1.5$~dex scatter in Li abundance---even at
fixed effective temperature---in \citeauthor{del15}'s
(\citeyear{del15}) survey of lithium in open clusters.  Surveys for Li
enhancement among metal-poor stars can be easier to interpret.
Inspired by \citeauthor{kra99}'s (\citeyear{kra99}) discovery of a
Li-rich giant in a metal-poor GC, \citet{pil00} surveyed 261 giants in
four metal-poor GCs, but they found no Li-rich giants.  Therefore, the
frequency of Li-rich red giants in GCs is less than 0.4\%.
\citet{dor14,dor15b} also surveyed red giants in GCs and found one
Li-rich giant out of about 350 giants, corresponding to a Li-rich
frequency of $(0.3 \pm 0.3)\%$.  \citet{ruc11} searched for Li-rich
giants in the Milky Way (MW) halo in the Radial Velocity Experiment
\citep[RAVE,][]{ste06}.  They found eight Li-rich giants out of 700
metal-poor field giants.  They also found one Li-rich giant in the GC
M68.  \citet{dom04} and \citet{kir12} also found 15 Li-rich giants in
MW dwarf satellite galaxies.  However, the MW field and dwarf galaxies
are not amenable to easily distinguishing between the AGB and upper
RGB\@.  In fact, many of the Li-rich giants discovered by
\citet{ruc11} and \citet{kir12} could be AGB stars.

Explanations for Li-rich RGB stars fall into three categories:
engulfment of a substellar companion, self-generation, and mass
transfer.  In the engulfment scenario
\citep[e.g.,][]{sie99,den00,mel05,vil09,ada12}, a red giant expands
into the orbit of a rocky planet, a hot Jupiter, or a companion brown
dwarf.  The destroyed companion could potentially enrich the star with
Li and other volatile elements that concentrate in planets
\citep{car13}.  Even if the engulfed companion does not donate Li to
its host star, it would provide angular momentum.  The resulting
increase in rotation rate could itself inspire deep mixing that
activates the Cameron--Fowler conveyor \citep{den04}.

In the self-generation scenario, stars can experience deep mixing
events that dredge Li to the stellar surface, where it is observable.
Rotationally induced mixing is one example.  Indeed, some Li-rich
giants are rapid rotators \citep{dra02,gui09,car10}, but others are
not \citep{ruc11}.  Other possible causes are mixing at the RGB
luminosity function bump \citep{cha00} or deep mixing inspired by He
core flashes at the tip of the RGB or on the HB
\citep{kum11,sil14,mon14}.  For example, \citet{dor15a} found a
Li-rich giant in the GC NGC~362 that may be either at the RGB bump
(hydrogen shell burning) or on the red clump (helium core burning).
On the other hand, \citeauthor{ant13}'s (\citeyear{ant13}) Li-rich
giant in NGC~6819 is one counter-example below the RGB bump.  The
chemical analysis of that star by \citet{car15} found no evidence for
deep mixing in any element other than Li.  Furthermore, most deep
mixing scenarios predict that the Li-rich giants would cluster at a
specific evolutionary phase (luminosity).  However, \citet{leb12}
found no luminosity clustering of Li-rich red giants.

Finally, stars can alter their surface compositions through binary
mass transfer.  AGB stars are known to generate carbon and
neutron-capture elements, like barium \citep{bus95}.  Hence, binary
companions to AGB stars or former AGB stars can be enhanced in those
elements \citep{mcc80}.  Intermediate-mass AGB stars can also dredge
up Li in the Cameron--Fowler conveyor.  Even less massive AGB stars
might be able to generate Li with the help of thermohaline mixing
\citep{can10}.  If the star transferred mass to a companion during a
phase of Li dredge-up, then that companion would be enhanced in Li.
This is a possible explanation for a Li-rich turn-off star in the GC
NGC~6397 \citep{koc11,pas14}.  That star will remain enhanced in Li
until the first dredge-up.  Assuming that the dredge-up dilutes a
fixed percentage of Li for all stars of similar mass and composition,
then the star would still appear Li-enhanced relative to other
post-dredge-up stars in the cluster.

GCs are the best environments to study low-mass stellar evolution.
The common distance to all the member stars makes it easy to determine
stellar luminosity.  The common age and small abundance dispersion in
most elements implies a similar evolution for all stars.  To first
order, a GC is a snapshot of stellar evolution over a sequence of
stellar masses at fixed age and mostly fixed metallicity.  With
reasonably attainable photometric uncertainty, the AGB and RGB can be
distinguished with a color--magnitude diagram (CMD) except for the
brightest giants, where the AGB nearly merges with the RGB\@.

We exploited the controlled stellar populations of GCs to study the
phenomenon of Li-rich giants.  We searched for Li-rich giants and
classified them photometrically as RGB or AGB\@.
Section~\ref{sec:obs} describes our observations, and
Section~\ref{sec:meas} details the measurement of Li and other
spectroscopic properties.  In Section~\ref{sec:li}, we define what it
means to be ``Li-rich'' and quantify the statistics of Li-rich giants
in GCs.  We address the possible origins of Li enhancement in
Section~\ref{sec:discussion}, and we summarize our conclusions in
Section~\ref{sec:summary}.


\section{Spectroscopic Observations}
\label{sec:obs}

\begin{deluxetable*}{lccl}
\tablewidth{0pt}
\tablecolumns{4}
\tablecaption{Globular Clusters Observed\label{tab:phot}}
\tablehead{\colhead{GC} & \colhead{RA (J2000)} & \colhead{Dec (J2000)} & \colhead{Source Catalogs}}
\startdata
NGC 288         & 00 52 45 & $-$26 34 57 & Stetson; \citet{bel01} \\
Pal 2           & 04 46 05 & $+$31 22 53 & Stetson \\
NGC 1904 (M79)  & 05 24 11 & $-$24 31 28 & Stetson; \citet{ros00} \\
NGC 2419        & 07 38 08 & $+$38 52 56 & \citet{ste00} \\
NGC 4590 (M68)  & 12 39 27 & $-$26 44 38 & Stetson; \citet{wal94} \\
NGC 5024 (M53)  & 13 12 55 & $+$18 10 05 & Stetson; \citet{an08} \\
NGC 5053        & 13 16 27 & $+$17 42 00 & Stetson; \citet{an08} \\
NGC 5272 (M3)   & 13 42 11 & $+$28 22 38 & \citet{ste00} \\
NGC 5634        & 14 29 37 & $-$05 58 35 & Stetson; \citet{bel02} \\
NGC 5897        & 15 17 24 & $-$21 00 36 & Stetson; \citet{tes01} \\
NGC 5904 (M5)   & 15 18 33 & $+$02 04 51 & \citet{ste00}; \citet{an08} \\
Pal 14          & 16 11 00 & $+$14 57 27 & \citet{sah05} \\
NGC 6205 (M13)  & 16 41 41 & $+$36 27 35 & Stetson \\
NGC 6229        & 16 46 58 & $+$47 31 39 & SDSS \\
NGC 6341 (M92)  & 17 17 07 & $+$43 08 09 & \citet{ste00}; \citet{an08} \\
NGC 6656 (M22)  & 18 36 23 & $-$23 54 17 & Stetson; \citet{pet94} \\
NGC 6779 (M56)  & 19 16 35 & $+$30 11 00 & \citet{hat04} \\
NGC 6838 (M71)  & 19 53 46 & $+$18 46 45 & Stetson \\
NGC 6864 (M75)  & 20 06 04 & $-$21 55 16 & \citet{kra07} \\
NGC 7006        & 21 01 29 & $+$16 11 14 & \citet{ste00}; \citet{an08} \\
NGC 7078 (M15)  & 21 29 58 & $+$12 10 01 & \citet{ste94,ste00} \\
NGC 7089 (M2)   & 21 33 27 & $-$00 49 23 & \citet{ste00}; \citet{an08} \\
NGC 7099 (M30)  & 21 40 22 & $-$23 10 47 & Stetson; \citet{san99} \\
Pal 13          & 23 06 44 & $+$12 46 19 & Stetson \\
NGC 7492        & 23 08 26 & $-$15 36 41 & Stetson \\
\enddata
\tablerefs{Cluster coordinates are from the compilation of \protect \citet[][updated 2010]{har96} and references therein.  ``Stetson'' refers to photometry by P.B.~Stetson.  Most of the photometry is available at \url{http://www2.cadc-ccda.hia-iha.nrc-cnrc.gc.ca/community/STETSON/standards/}, but Stetson provided some of it directly to us.  ``SDSS'' refers to photometry from the Sloan Digital Sky Survey \citep{aba09}.}
\end{deluxetable*}

We observed \ncluster~GCs with Keck/DEIMOS \citep{fab03} over eight
years.  Table~\ref{tab:phot} lists the clusters and their coordinates.
Some of these slitmasks were observed with the purpose of validating a
method to measure metallicities and $\alpha$ element abundances from
DEIMOS spectra.  \citet{kir08a,kir10} previously published these
observations.  Most of the remaining slitmasks were designed expressly
to search for Li-rich red giants.

\subsection{Source Catalogs}

We used custom slitmasks designed to observe giant stars in the
clusters.  In order to design the slitmasks, we used photometric
catalogs from various sources.

Our primary source of photometry was P.B.~Stetson's database of
photometric standard fields.  We downloaded some of these from
Stetson's public web page, but he provided some of these catalogs to
us privately \citep[see][]{kir10}.  Several of these clusters were
also previously published \citep{ste94,ste00}.  These catalogs were
made with DAOPHOT \citep{ste87,ste11}, which models the point spread
functions (PSFs) of stars.  This approach performs better than
aperture photometry in crowded fields, like GCs.

Some of Stetson's clusters had dense sampling over a field comparable
in size to a DEIMOS slitmask.  In these cases, we relied on his
photometry alone.  The catalogs for other clusters sampled only tens
or hundreds of stars for the purposes of providing a photometric
calibration field rather than a science catalog.  In these cases, we
supplemented Stetson's photometry with other sources.
Table~\ref{tab:phot} lists the source catalogs for each cluster.
Notable sources include the Sloan Digital Sky Survey
\citep[SDSS,][]{aba09} and \citet{an08}.  Because the primary SDSS
catalog uses aperture photometry, \citet{an08} re-reduced the
photometry of select GCs with DAOPHOT\@.

All of the catalogs used have coverage in at least two of the three
filters $B$, $V$, and $I$.  We corrected the observed magnitudes for
extinction according to the dust maps of \citet{sch98}.

\subsection{Target Selection}

We designed the slitmasks with a minimum slit length of $4\arcsec$ and
separation between slits of $0\farcs 35$.  These choices allowed just
enough separation between stars to (1) avoid overlapping spectra and
(2) permit sky subtraction from the empty portions of the slits.
However, these restrictions also forced us to choose among the many
stars in the dense GCs.  Although several hundred GC giants might have
been visible in a single DEIMOS pointing, the slitmask would allow
only about 150 targets at most.  We developed target selection
strategies to pick out likely giant members of the GCs.

Because the \nmask~slitmasks were designed for different projects over
many years, the target selection strategy was not uniform.  Although
most masks were designed for giants, some included main sequence
stars.  In general, selection along the RGB was performed by defining
selection regions in the CMD\@.  In some cases, where the RGB was
well-defined and distinct from the foreground, we drew an irregular
polygon around the RGB and selected stars inside of it.  In other
cases, we drew an old ($\sim 12$~Gyr) isochrone corresponding to the
metallicity of the cluster \citep[][updated 2010]{har96}.  We used
both Victoria--Regina \citep{van06} and Yonsei--Yale \citep{dem04}
isochrone models.  The selection region was defined within a color
range (typically 0.1~mag) around the isochrone.  For most slitmasks,
brighter stars were given higher priority for selection.

The target selection favored first-ascent RGB stars rather than
helium-burning stars on the HB or AGB\@.  The HB was particularly
disfavored because the spectra of hot, blue stars do not readily lend
themselves to the measurement of radial velocity and metallicity,
which was the original intent for many of the slitmasks.  Therefore,
this data set is not ideal to search for Li-richness on the HB\@.
However, it is suitable for quantifying the frequency of Li-rich
giants on the RGB or upper AGB\@.

\begin{figure*}[p!]
\centering
\includegraphics[width=\textwidth]{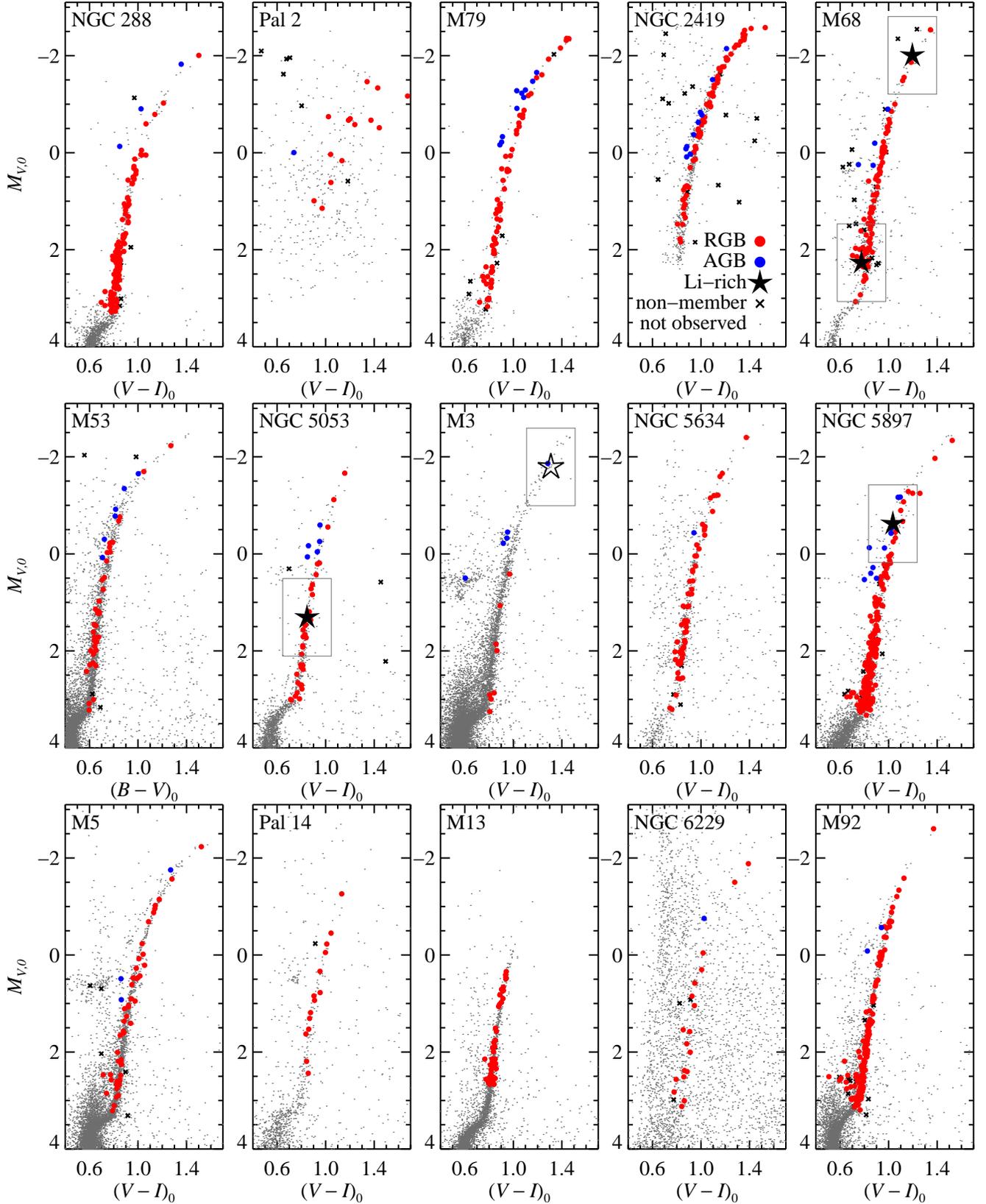}
\caption{Color--magnitude diagrams for all \ncluster\ GCs observed
  with DEIMOS\@.  The panel for NGC~2419 includes a figure legend.
  Li-rich stars are shown as black, five-pointed stars.  The hollow
  star indicates the Li-rich giant IV--101 in M3 \citep{kra99}, which
  is not part of our sample.  Spectroscopically confirmed, Li-normal
  members are shown as red (RGB) and blue (AGB) points.  Non-members
  are shown as black crosses.  Gray points show stars that we did not
  observe with DEIMOS\@.  We distinguished between RGB and AGB stars
  by drawing selection regions in the CMDs.  Figure~\ref{fig:cmdli}
  shows detail of the gray boxes around the Li-rich
  stars.\label{fig:cmd}}
\end{figure*}

\addtocounter{figure}{-1}
\begin{figure*}[t!]
\centering
\includegraphics[width=\textwidth]{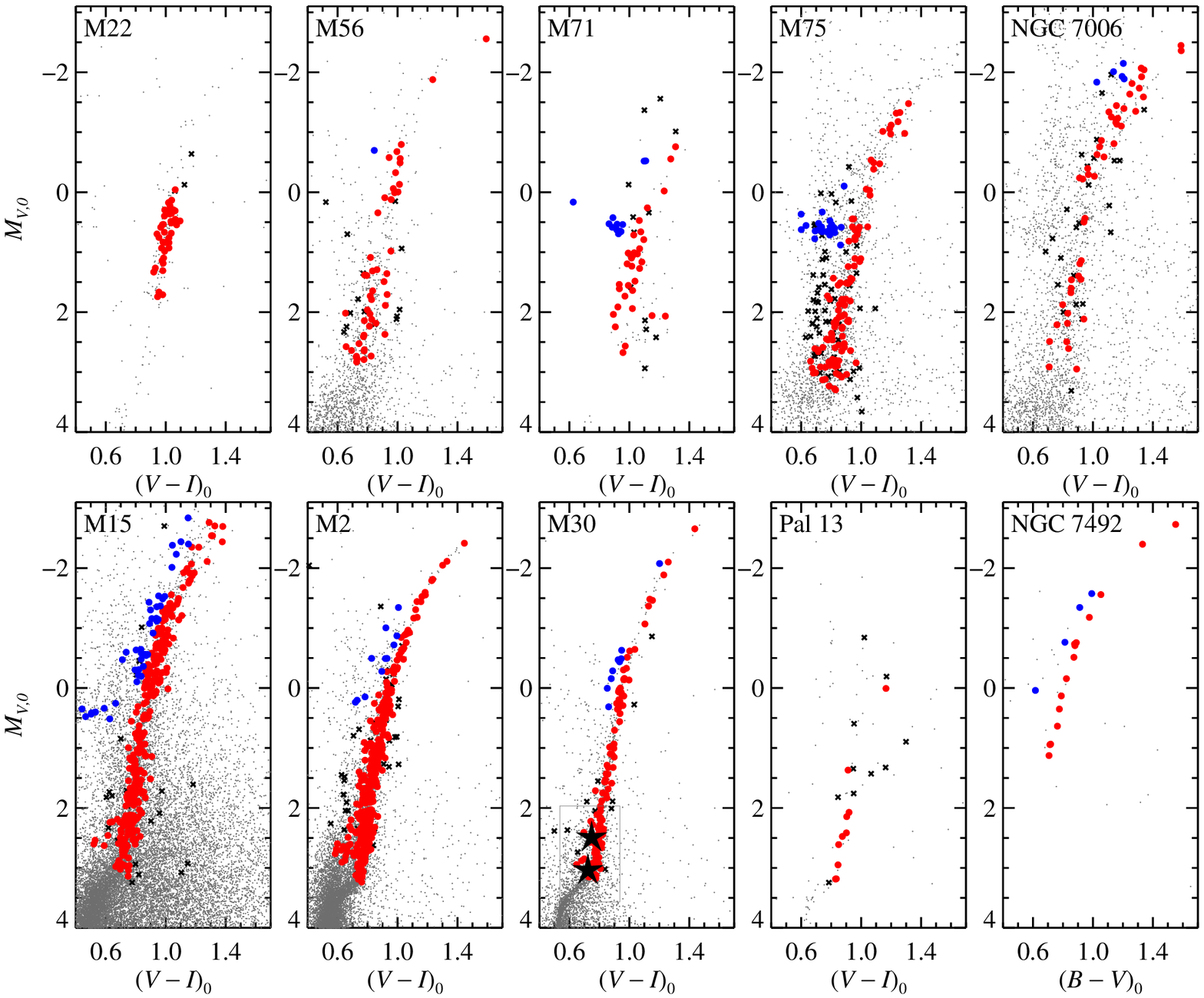}
\caption{\emph{--- continued ---}}
\end{figure*}

Figure~\ref{fig:cmd} shows the extinction- and reddening-corrected
CMDs for all of the GCs in our sample.  M53 and NGC~7492 are shown
with $(B-V)_0$ color, and all of the other GCs are shown with
$(V-I)_0$.  Stars that we identified as members
(Section~\ref{sec:membership}) are shown as colored points or black,
five-pointed stars.

\subsection{Separation of RGB and AGB}

GCs are excellent laboratories to study stellar evolution because they
are nearly single-age populations of nearly uniform
metallicity.\footnote{There is extensive observational evidence that
  GCs are chemically complex \citep[e.g.,][]{gra04,gra12}.  In
  particular, primordial intracluster variation in certain elements,
  such as O, Na, Mg, and Al, indicates that cluster stars were
  differentially enhanced with the products of high-temperature
  hydrogen burning.  Li is not immune to the primordial variation, as
  exhibited by weak Li--Na and Li--Al anti-correlations observed in
  some clusters \citep{lin09b,mon12,dor15b}.  Unfortunately, we cannot
  distinguish between first and later generation stars in our sample
  because we cannot observe Na in our spectra.  However, GCs are still
  simple enough for our purposes.  Specifically, it is straightforward
  to distinguish the AGB from the RGB, and the heavy elements, like
  Fe, are invariant within each of the clusters in our sample.}  For
these reasons, GCs are the best stellar populations for distinguishing
between the AGB and the RGB\@.  This distinction helps determine how
evolutionary phase plays a role in Li-richness.

Although model isochrones could be used for this task, we found that
small imperfections in the models resulted in misidentification at
high stellar luminosities, where the AGB asymptotically approaches the
RGB\@.  Instead, we identified the AGB ``by eye.''  We drew a
selection region around the AGB for each GC\@.  AGB stars are shown in
blue in Figure~\ref{fig:cmd}.  RGB stars are shown in red.

\subsection{Observations}

\addtocounter{table}{1}

Table~\ref{tab:obs} lists the observing log, including the slitmask
name, the number of targets on the slitmask, the date of observation,
the airmass and seeing at the time of observation, the number of
exposures, and the total exposure time.  The number of targets is the
number of science slitlets in the mask (excluding alignment boxes).
It is not the number of stars in the final sample.  In addition to
member giants, the slitmasks included main sequence stars as well as
non-members.

All slitmasks were observed with the 1200G grating, which has a groove
spacing of 1200~mm$^{-1}$ and a blaze wavelength of 7760~\AA\@.  The
slit widths were typically $0\farcs 7$.  The resulting resolution was
1.2~\AA, which corresponds to a resolving power of $R \approx 6500$ at
the blaze wavelength.  Each pixel encompasses 0.33~\AA, such that a
resolution element spans 3.6~pixels.  Slitmasks with the letter ``l''
were observed at a central wavelength of 7500~\AA\@.  Other slitmasks
were observed at a central wavelength of 7800~\AA\@.  The OG550
order-blocking filter blocked second- and higher-order light from
contaminating the spectra.  We used DEIMOS's flexure compensation
system, which provides wavelength stability of about 0.03~\AA\ during
the observation of one slitmask.  Afternoon calibrations included
exposures of a quartz lamp for flat fielding and an exposure of Ne,
Kr, Ar, and Xe arc lamps for wavelength calibration.

We reduced the DEIMOS spectra with the \texttt{spec2d} IDL data
reduction pipeline developed by the DEEP2 team \citep{coo12,new13}.
The pipeline excises the 2-D spectrum for each slitlet.  The 2-D
spectrum is flat-fielded and wavelength-calibrated.  The wavelength
calibration from the arc lamps is refined with night sky emission
lines.  All of the exposures are combined, and cosmic rays are
removed.  Finally, the 1-D spectrum is extracted with optimal
extraction.  The software tracks the variance spectrum at every step.
The result is a flat-fielded, wavelength-calibrated, 1-D spectrum of
the target along with an estimate of the error in each pixel.

\begin{figure}[t!]
\centering
\includegraphics[width=\columnwidth]{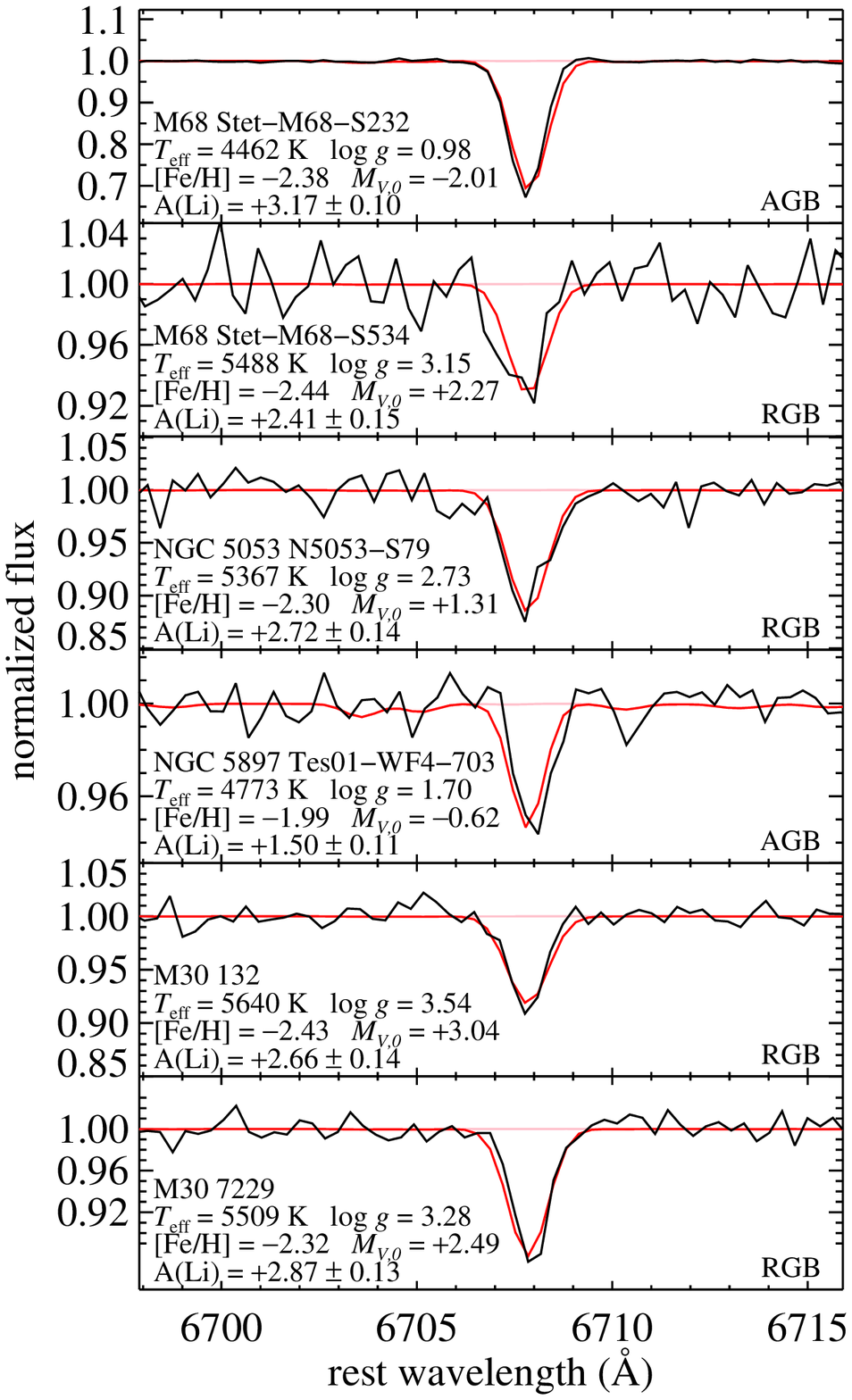}
\caption{DEIMOS spectra (black) of the six Li-rich giants around the
  \liin\ absorption line.  Best-fit synthetic spectra are shown in
  red.  The pink lines show synthetic spectra with no Li.  Each panel
  gives the star's host cluster, the star's name, temperature,
  gravity, metallicity, luminosity, and NLTE-corrected Li
  abundance.\label{fig:spectra}}
\end{figure}

Figure~\ref{fig:spectra} shows the spectra of the six giants that we
determined to be Li-rich members of their respective GCs (see
Section~\ref{sec:li}).  Only a small spectral region around the
\liin\ line is shown.


\section{Spectroscopic Measurements}
\label{sec:meas}

We measured four important parameters from each spectrum: radial
velocity, $v_{\rm helio}$; effective temperature, $T_{\rm eff}$; metallicity,
[Fe/H]; and Li abundance, $A({\rm Li})$.

\subsection{Radial Velocities}

We measured $v_{\rm helio}$ in the same manner as \citet{sim07}.
First, we measured $v_{\rm obs}$, the velocity required to shift the
spectrum into the rest frame.  To do so, we cross-correlated each
spectrum with 16 template spectra observed with DEIMOS, kindly
provided by \citeauthor{sim07}.  Because of imperfect centering in the
slitlet, a star can have an apparent radial velocity with respect to
the telluric absorption lines.  To counteract this error, we
cross-correlated each spectrum with a template spectrum of a hot star,
which is dominated by telluric absorption.  The resulting velocity is
$v_{\rm center}$, the velocity required to shift the spectrum into the
geocentric frame.  We also computed $v_{\rm corr}$, the correction
required to shift from the geocentric to the heliocentric frame.  The
final heliocentric velocity of the star is $v_{\rm helio} = v_{\rm
  obs} + v_{\rm center} + v_{\rm corr}$.

\subsection{Atmospheric Parameters}
\label{sec:atm}

We measured $T_{\rm eff}$ and [Fe/H] in the same manner as
\citet{kir08a,kir10}.  This section summarizes the procedure.  First,
we shifted the spectrum into the rest frame, removed telluric
absorption by dividing by the spectrum of a hot star, and divided by
the continuum, approximated by a spline with a breakpoint spacing of
100~\AA\@.  Next, we searched for the best-fitting synthetic spectrum
among a large grid of spectra computed with MOOG \citep{sne73} and
ATLAS9 model atmospheres \citep{kur93,sbo05}.

We estimated initial guesses at $T_{\rm eff}$ and surface gravity,
$\log g$, by comparing the star's color and magnitude to model
isochrones shifted by the distance modulus of its respective GC\@.  In
searching the grid, $T_{\rm eff}$ was allowed to vary in a range
around the photometrically determined value, but $\log g$ was fixed at
the photometric value.  On the other hand, no restrictions were
imposed on [Fe/H]\@.

We made ``first draft'' measurements of $T_{\rm eff}$ and [Fe/H] by
minimizing $\chi^2$ between the observed and synthetic spectra.  We
refined these measurements by using the best-fit synthetic spectrum to
improve the continuum determination.  We repeated this iterative
continuum refinement until $T_{\rm eff}$ and [Fe/H] changed by a
negligible amount between iterations.  The values of $T_{\rm eff}$ and
$\log g$ at the end of the last iteration were regarded as the final
measurements.

\subsection{Membership}
\label{sec:membership}

We considered only stars that are members of our sample of GCs for the
purposes of this project.  Our measurements of atmospheric parameters
are valid only for member stars because we used model isochrones to
estimate $T_{\rm eff}$ and $\log g$.  The measurements are not valid
for mon-member stars at unknown distances.

First, we removed duplicate spectra.  Where a star was observed
multiple times on different slitmasks, we kept the measurement with
the lowest estimate of error on [Fe/H], which is essentially a S/N
criterion.  We removed \ndup~duplicate spectra.

Second, we eliminated any stars that were obviously non-members or
non-giants based on their positions in the CMD\@.  Although the
slitmasks were designed to avoid non-members, some obvious non-members
were placed on the slitmask merely to fill it with targets.  We drew a
generous CMD selection region around the stellar locus and flagged
stars outside of the region as non-members.  Figure~\ref{fig:cmd}
shows some of these non-members as crosses.  We also eliminated
non-giant stars by imposing a cut on surface gravity: $\log g < 3.6$.

Third, we restricted the member list on the basis of radial velocity.
We estimated the cluster's mean velocity, $\langle v_{\rm helio}
\rangle$, and velocity dispersion, $\sigma_v$, by calculating the mean
velocity of all stars within 40~km~s$^{-1}$ of the median velocity.
We compiled a list of all stars that satisfied $|v_{\rm helio} -
\langle v_{\rm helio} \rangle| < 2.58\sigma_v$ (99\% of all stars in a
Gaussian velocity distribution).  From this list, we re-computed
$\langle v_{\rm helio} \rangle$ and $\sigma_v$.  The member list
includes only those stars that have $|v_{\rm helio} - \langle v_{\rm
  helio} \rangle| - \delta v < 3\sigma_v$, where $\delta v$ is the
uncertainty on the radial velocity.  In other words, any star whose
$1\sigma$ velocity error bar overlapped the $3\sigma_v$ membership cut
was allowed as a member.  Although the different criteria for stars
used in the computation of $\sigma_v$ versus the member list may seem
capricious, we found from examining the velocity histograms that this
procedure reliably identified stars in the GC's velocity peak.

Finally, we restricted the member list on the basis of [Fe/H]\@.  The
procedure was nearly identical to the velocity membership criterion.
The mean metallicity, $\langle {\rm [Fe/H]} \rangle$, and metallicity
dispersion, $\sigma({\rm [Fe/H]})$, were computed from all stars in
the cluster.  Then, these values were re-computed from a more
restricted list: $|{\rm [Fe/H]} - \langle {\rm [Fe/H]} \rangle| <
2.58\sigma({\rm [Fe/H]})$ and ${\rm [Fe/H]} < -0.5$.  With these
refined values, the final membership list was those stars with $|{\rm
  [Fe/H]} - \langle {\rm [Fe/H]} \rangle| - \delta{\rm [Fe/H]} <
3\sigma({\rm [Fe/H]})$, where $\delta{\rm [Fe/H]}$ is the uncertainty
on [Fe/H].

\subsection{Li Abundance}
\label{sec:li_measurement}

\begin{deluxetable}{llcc}
\tablewidth{0pt}
\tablecolumns{4}
\tablecaption{Line List\label{tab:linelist}}
\tablehead{\colhead{Wavelength (\AA)} & \colhead{Species} & \colhead{EP (eV)} & \colhead{$\log(gf)$}}
\startdata
\nodata   & \nodata       & \nodata & \nodata \\
6707.752  & Sc$\,${\sc i} & 4.049 & $-$2.672\phn  \\
6707.7561 & $^7$Li$\,${\sc i} & 0.000 & $-$0.4283 \\
6707.7682 & $^7$Li$\,${\sc i} & 0.000 & $-$0.2062 \\
6707.771  & Ca$\,${\sc i} & 5.796 & $-$4.015\phn  \\
6707.799  & CN            & 1.206 & $-$1.967\phn  \\
6707.9066 & $^7$Li$\,${\sc i} & 0.000 & $-$1.5086 \\
6707.9080 & $^7$Li$\,${\sc i} & 0.000 & $-$0.8069 \\
6707.9187 & $^7$Li$\,${\sc i} & 0.000 & $-$0.8069 \\
6707.9196 & $^6$Li$\,${\sc i} & 0.000 & $-$0.4789 \\
6707.9200 & $^7$Li$\,${\sc i} & 0.000 & $-$0.8069 \\
\nodata   & \nodata       & \nodata & \nodata \\
\enddata
\tablerefs{Lithium lines are from \citet{hob99}.  Other lines are from \citet{kir08a}, which is compilation of atomic lines from VALD \citep{kup99} and molecular lines from \citet{kur93}.}
\tablecomments{Wavelengths are in air.  (This table is available in its entirety in a machine-readable form in the online journal.  A portion is shown here for guidance regarding its form and content.)}\end{deluxetable}

We measured Li abundances by spectral synthesis of the \liin\ doublet.
We compiled a line list (Table~\ref{tab:linelist}) of absorption lines
in the region 6697--6717~\AA\@.  The Li absorption lines come from
\citeauthor{hob99}'s (\citeyear{hob99}) list.  Other lines are from
\citeauthor{kir08a}'s (\citeyear{kir08a}) compilation from VALD
\citep[for neutral and ionized atoms,][]{kup99} and \citet[for
  molecules,][]{kur93}.  The Li lines are separated by isotope ($^6$Li
and $^7$Li).

We prepared the spectrum by performing a local continuum correction
around \liin\@.  We used MOOG and \citeauthor{kir11}'s
(\citeyear{kir11}) grid of ATLAS9 model atmospheres to compute a
synthetic spectrum devoid of Li.  The atmospheric parameters ($T_{\rm
  eff}$, $\log g$, [Fe/H]) were tailored to each star following the
procedure in Section~\ref{sec:atm}.  The microturbulent velocity
($\xi$) was calculated based on a calibration between $\xi$ and $\log
g$ \citep{kir09}.  We divided the observed spectrum by this model.  We
fit a straight line with variable slope and intercept to the residual
in the wavelength range 6697--6717~\AA, but excluding the Li doublet
(6705.7--6709.9~\AA)\@.  This linear fit comprised the local continuum
correction, by which we divided the observed spectrum.

We measured $A({\rm Li})$ in the observed spectrum by minimizing
$\chi^2$ between the continuum-refined, observed spectrum and a model
spectrum.  The only free parameter in the fit was $A({\rm Li})$.  We
minimized $\chi^2$ with the Levenberg--Marquardt IDL code
\texttt{MPFIT} \citep{mar12}.  This required computing many spectral
syntheses with MOOG, which we did in the same manner as for the
Li-free spectrum described in the previous paragraph.

We set the $^7$Li/$^6$Li isotopic ratio to 30.  Although
\liin\ spectra modeled with 3D, NLTE model atmospheres show no
detectable $^6$Li \citep{lin13}, a $^7$Li/$^6$Li ratio of $\sim
30$---while not an accurate representation of the atmospheric
composition---gives the best-fitting line shape in a 1D, LTE spectral
synthesis \citep{smi98}, such as ours.  Our results are nearly
insensitive to this choice because the resolution of our spectra is
smaller than the isotopic splitting of the Li doublet.

We took the $1\sigma$ error on $A({\rm Li})$ to be the value by which
$A({\rm Li})$ needed to change in order to raise $\chi^2$ by 1 from
the minimum $\chi^2$.  For spectra with ${\rm S/N} >
300$~pixel$^{-1}$, this estimate of the error could be even smaller
than 0.01, which is unrealistically low because it does not account
for systematic error, such as imperfections in the spectral model.  We
imposed a minimum error of 0.1~dex by adding 0.1~dex in quadrature
with the statistical error.

Most of the stars had no detectable Li.  For these stars, the $\chi^2$
contour flattened to a constant value at low $A({\rm Li})$.  We
computed $2\sigma$ upper limits as the value of $A({\rm Li})$
corresponding to an increase in $\chi^2$ of 4 above the minimum
$\chi^2$.  We found this value using a truncated Newton minimization
method.\footnote{\texttt{TNMIN}, an IDL code by C.~Markwardt
  (\url{http://cow.physics.wisc.edu/~craigm/idl/idl.html}).}

We examined the spectrum of every Li doublet to confirm that the
measurement of $A({\rm Li})$ or its upper limit is valid.  We plotted
the best-fitting synthetic spectrum over the continuum-corrected
observed spectrum.  If the fit appeared to fail, then we removed the
spectrum from our sample.  Common reasons for failure included
single-pixel noise spikes (possibly due to cosmic rays) and badly
placed continuum measurements due to spectral artifacts.  We also
flagged every spectrum with a convincing detection of Li.  Although we
technically measured $A({\rm Li})$ for every spectrum, we present
upper limits for those spectra with unconvincing detections.

\citet{lin09a} computed corrections to $A({\rm Li})$ to counteract
deficiencies from the assumption of local thermodynamic equilibrium
(LTE) in computing synthetic spectra.  The non-LTE (NLTE) correction
depends on the LTE lithium abundance and stellar parameters, like
$T_{\rm eff}$ and $\log g$.  \citeauthor{lin09a}\ provided convenient
tables to compute NLTE corrections for most cool stars.  All of the
values of $A({\rm Li})$, including upper limits, in the text, figures,
and tables in this paper have these NLTE corrections applied.  We
linearly extrapolated the correction for stars with stellar parameters
outside of the range of \citet{lin09a}'s tables.

\addtocounter{table}{1}

Table~\ref{tab:catalog} gives Li measurements or $2\sigma$ upper
limits for our sample.  The table also identifies whether the star
fell in the RGB or AGB selection window.  Non-members and stars that
were removed from the sample upon visual inspection are not shown in
the table.  The table gives the six Li-rich giants first, followed by
all other stars in order of right ascension.  The photometric
magnitudes and colors are corrected for extinction and reddening.


\section{Li Enhancement}
\label{sec:li}

In this section, we quantify the number of Li-rich giants in our
sample.  To do so, we establish a quantitative definition for
``Li-richness.''  We also separate the statistics on Li-richness by
stellar evolutionary state (RGB or AGB)\@.

\subsection{Defining ``Li-Rich''}

\begin{figure*}[t!]
\centering
\includegraphics[width=0.7\textwidth]{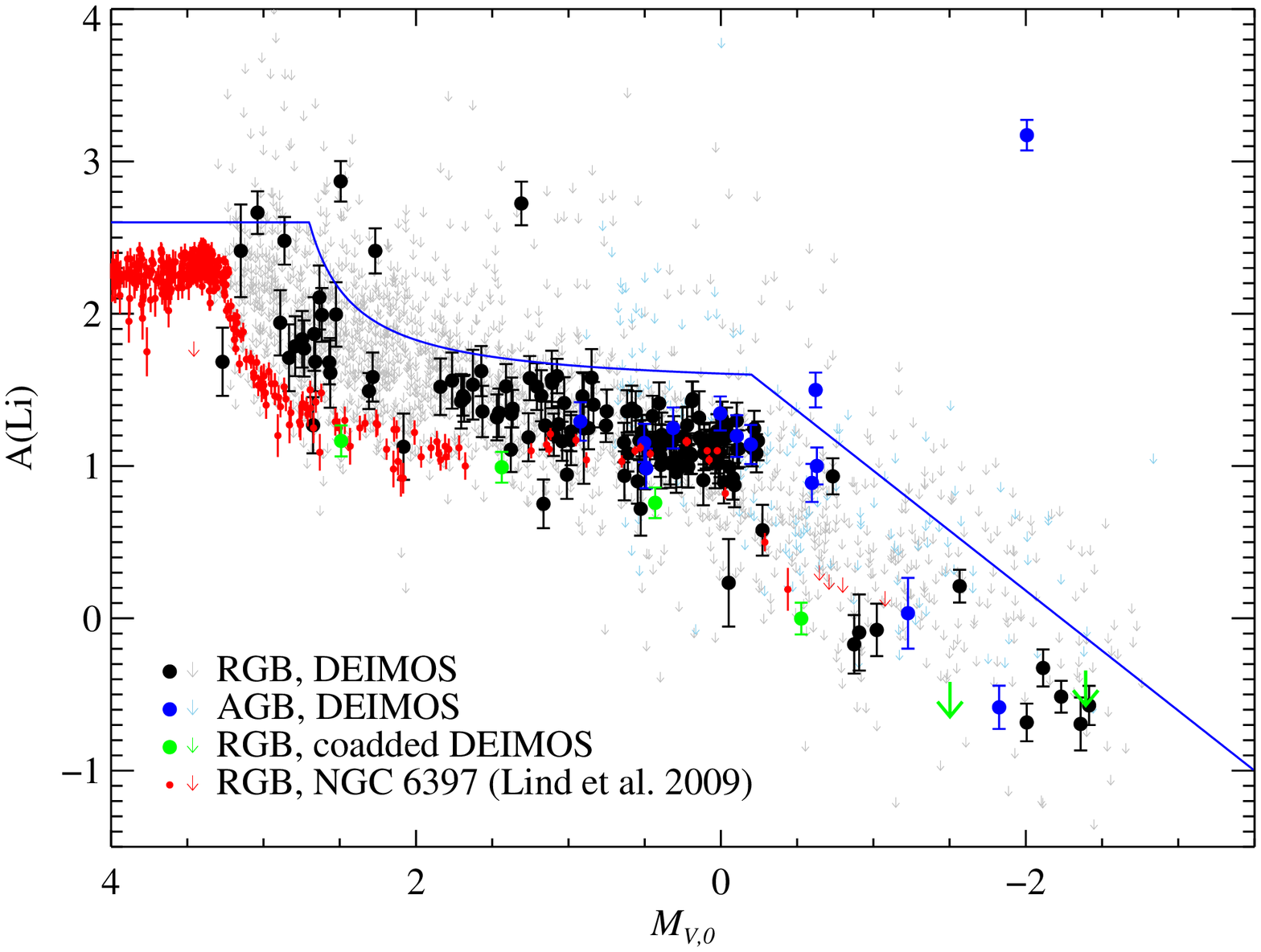}
\caption{NLTE-corrected Li abundances versus absolute magnitude.  Our
  DEIMOS detections of Li are shown as black (RGB) and blue (AGB)
  points.  Upper limits are shown in gray (RGB) and faded blue
  (AGB)\@.  For comparison, high-resolution spectroscopic measurements
  of Li in the GC NGC~6397 \citep{lin09b} and DEIMOS spectra of red
  giants coadded in bins of $M_{V,0}$ are shown in red and green,
  respectively.  The blue curve (Equation~\ref{eq:lirich}) separates
  Li-rich from Li-normal stars.\label{fig:alimv}}
\end{figure*}

In order to define ``Li-rich,'' we examine what it means to be
``Li-normal.''  The definition should depend on the star's luminosity
because surface Li is progressively depleted as the star ascends the
RGB\@.  For example, a giant with $A({\rm Li}) = 1.1$ and $M_V = +1$
would not be Li-rich, but stars with $M_V < 0$ begin a phase of Li
destruction at the luminosity function bump (in contrast with
\textit{dilution} at the first dredge-up).  As a result, a star with
$A({\rm Li}) = 1.1$ and $M_V = -2$ would be Li-rich.

\citet{lin09b} conducted the definitive study of Li in GC stars.  They
measured $A({\rm Li})$ for hundreds of stars in the metal-poor GC
NGC~6397 from $R = 14,000$, high-S/N VLT/FLAMES spectroscopy.
Figure~\ref{fig:alimv} shows their measurements in red.  The main
sequence stars at $M_V > +3.3$ have a constant $A({\rm Li}) = 2.3$.
The first dredge-up begins at $M_V = +3.3$ and depletes $A({\rm Li})$
to 1.1.  The Li remains briefly untouched until the luminosity
function bump at $M_V = 0.0$, which further depletes Li to an
undetectable level.

Our DEIMOS spectra have lower spectral resolution than
\citeauthor{lin09b}'s FLAMES spectra.  Consequently, we did not detect
Li in the majority of our stars.  Those stars with detections also
have larger $A({\rm Li})$ uncertainties than the FLAMES measurements.
The DEIMOS detections tend to be for stars with larger $A({\rm Li})$
at fixed $M_V$ than the FLAMES detections.  That is why most of our
detections of Li trace the upper envelope of the NGC~6397 data.  We
have detected Li only in those stars with upward fluctuations in
$A({\rm Li})$ due to intrinsic variation in the cluster or, more
likely, random noise in the DEIMOS spectra.

We also quantified what it means to be Li-normal by coadding DEIMOS
spectra of RGB stars in six bins of $M_V$.  The least luminous bin was
$M_V > +2$, and the most luminous bin was $M_V < -2$.  The other four
bins were 1~mag wide in the range $+2 > M_V > -2$.  In each bin, we
coadded all RGB spectra (excluding the AGB) that do not satisfy the
Li-rich criterion (Equation~\ref{eq:lirich}).  We did not include
spectra that we identified in Section~\ref{sec:li_measurement} to be
problematic.  The spectra were interpolated onto a common wavelength
array and coadded with inverse variance weighting.  We also averaged
$M_V$ and atmospheric parameters in each bin, weighting by the median
inverse variance within 10~\AA\ of \liin\@.  We measured $A({\rm
  Li})$, treating the coadded spectrum as a single spectrum with a
single $T_{\rm eff}$, $\log g$, and [Fe/H], which were fixed at the
weighted average values for all the spectra in the bin.

\begin{figure}[t!]
\centering
\includegraphics[width=\columnwidth]{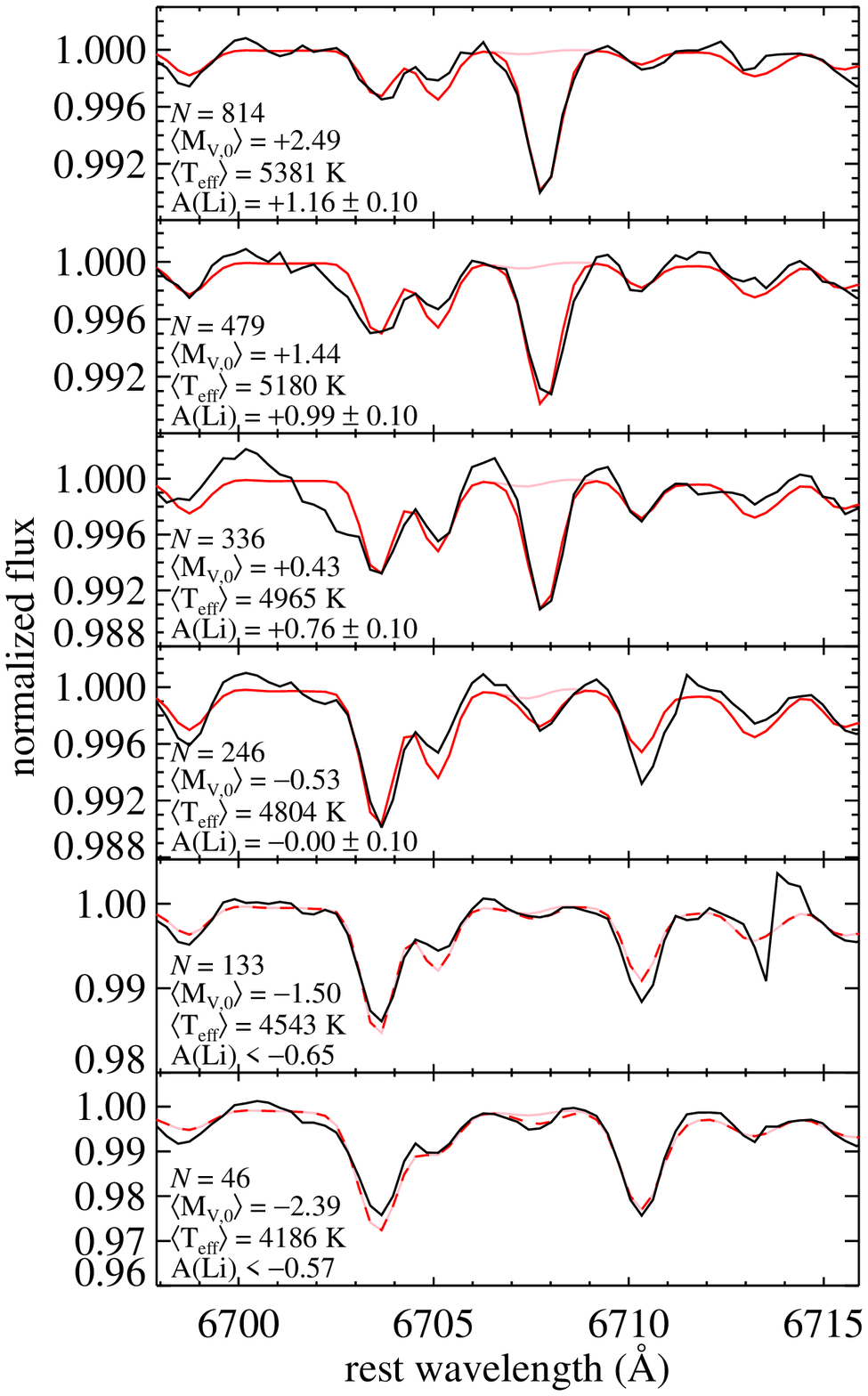}
\caption{Coadded spectra of RGB stars in bins of absolute magnitude
  ($M_{V,0}$), in order of least luminous (top) to most luminous
  (bottom).  Each panel shows the number of spectra in the coaddition
  as well as the average $M_{V,0}$, average $T_{\rm eff}$, and
  NLTE-corrected Li abundance.  Weak absorption lines not apparent in
  Figure~\ref{fig:spectra} are visible because the $y$-axis ranges of
  this figure are much smaller.  Best-fitting synthetic spectra are
  shown in red.  The red spectra in the bottom two panels show the
  spectra corresponding to the $2\sigma$ upper limit on $A({\rm Li})$
  (represented by the broken lines).  Synthetic spectra without Li are
  shown in pink.\label{fig:coadd}}
\end{figure}

Figure~\ref{fig:coadd} shows the coadded spectra.  Most absorption
lines become stronger with increasing luminosity (note the increasing
$y$-axis range) because $T_{\rm eff}$ decreases with increasing
luminosity on the RGB\@.  However, \liin\ becomes weaker because Li is
depleted with decreasing $T_{\rm eff}$.  Figure~\ref{fig:alimv}
compares our coaddition measurements of $\langle A({\rm Li}) \rangle$
(green) to individual stars in NGC~6397 (red).  Except for the
$\langle M_{V,0} \rangle = -0.4$ bin, the green points lie in the
midst of the red points.  The bin with $\langle M_{V,0} \rangle =
-0.4$ shows a likely spurious absorption feature at 6702~\AA, which
pushes up the continuum.  Therefore, our measurement of $A({\rm Li})$
might be slightly low in this bin.

The high-quality NGC~6397 data along with our coadded DEIMOS data
define a clear trend of $A({\rm Li})$ with $M_V$.  We drew a boundary
in Figure~\ref{fig:alimv} along the upper envelope of our
measurements.  The following equation defines the boundary:

\begin{eqnarray}
A({\rm Li}) = \left\{
\begin{array}{lcr}
2.60                          & ~ & M_V \ge +2.7 \\
1.50 + \frac{0.33}{3.0 - M_V} & ~ & +2.7 > M_V > -0.2 \\
1.76 + 0.79\,M_V              & ~ & M_V \le -0.2 \\
\end{array}
\right. \label{eq:lirich}
\end{eqnarray}

Six Li detections fall above the boundary.  Although the exact
placement of the boundary is somewhat subjective,
Figure~\ref{fig:alimv} shows that there is little ambiguity about
which stars are Li-rich.  The assignment of Li-rich and Li-normal
could be questioned only for the faintest Li-rich star, M30~132.  A
more rigorous analysis might use multiple levels of Li-richness, such
as ``Li-normal,'' ``Li-rich'', and ``super Li-rich,'' or even a
continuously defined ``Li-richness'' variable.  For simplicity, we
retain our binary (yes/no) definition, accepting that the Li-richness
of M30~132 is ambiguous.

\subsection{Li-Rich Frequency}

\begin{deluxetable*}{lccrrrrrrccc}
\tablewidth{0pt}
\tablecolumns{12}
\tablecaption{Li-Rich Statistics\label{tab:gc}}
\tablehead{\colhead{GC} & \colhead{$M_V$} & \colhead{[Fe/H]} & \multicolumn{2}{c}{Members\tablenotemark{a}} & \multicolumn{2}{c}{Li-Rich} & \multicolumn{2}{c}{Li-Normal\tablenotemark{b}} & \multicolumn{3}{c}{Li-Rich Frequency (\%)} \\
\colhead{ } & \colhead{ } & \colhead{ } & \colhead{RGB} & \colhead{AGB} & \colhead{RGB} & \colhead{AGB} & \colhead{RGB} & \colhead{AGB} & \colhead{RGB} & \colhead{AGB} & \colhead{Both}}
\startdata
NGC 288         & $-6.8$ & $-1.32$ &  106 &    3 &    0 &    0 &   97 &    3 &       $< 1.0$ &        $< 33$ &       $< 1.0$ \\
Pal 2           & $-8.0$ & $-1.42$ &   15 &    1 &    0 &    0 &    3 &    0 &        $< 33$ &       \nodata &        $< 33$ \\
NGC 1904 (M79)  & $-7.9$ & $-1.60$ &   68 &   10 &    0 &    0 &   64 &   10 &       $< 1.6$ &        $< 10$ &       $< 1.4$ \\
NGC 2419        & $-9.4$ & $-2.15$ &   73 &   10 &    0 &    0 &   18 &    1 &       $< 5.6$ &        $<100$ &       $< 5.3$ \\
NGC 4590 (M68)  & $-7.4$ & $-2.23$ &   87 &    5 &    1 &    1 &   80 &    4 & $1.2 \pm 1.2$ &   $20 \pm 20$ & $2.3 \pm 1.6$ \\
NGC 5024 (M53)  & $-8.7$ & $-2.10$ &   38 &    6 &    0 &    0 &   23 &    6 &       $< 4.3$ &        $< 17$ &       $< 3.4$ \\
NGC 5053        & $-6.8$ & $-2.27$ &   39 &    5 &    1 &    0 &   22 &    5 & $4.3 \pm 4.3$ &        $< 20$ & $3.6 \pm 3.6$ \\
NGC 5272 (M3)   & $-8.9$ & $-1.50$ &    8 &    5 &    0 &    0 &    8 &    4 &        $< 13$ &        $< 25$ &       $< 8.3$ \\
NGC 5634        & $-7.7$ & $-1.88$ &   60 &    1 &    0 &    0 &   42 &    1 &       $< 2.4$ &        $<100$ &       $< 2.3$ \\
NGC 5897        & $-7.2$ & $-1.90$ &  216 &   11 &    0 &    1 &  190 &   10 &       $< 0.5$ & $9.1 \pm 9.1$ & $0.5 \pm 0.5$ \\
NGC 5904 (M5)   & $-8.8$ & $-1.29$ &   48 &    3 &    0 &    0 &   48 &    3 &       $< 2.1$ &        $< 33$ &       $< 2.0$ \\
Pal 14          & $-4.8$ & $-1.62$ &   14 &    0 &    0 &    0 &    1 &    0 &        $<100$ &       \nodata &        $<100$ \\
NGC 6205 (M13)  & $-8.6$ & $-1.53$ &   55 &    0 &    0 &    0 &   51 &    0 &       $< 2.0$ &       \nodata &       $< 2.0$ \\
NGC 6229        & $-8.1$ & $-1.47$ &   18 &    1 &    0 &    0 &   13 &    1 &       $< 7.7$ &        $<100$ &       $< 7.1$ \\
NGC 6341 (M92)  & $-8.2$ & $-2.31$ &  149 &    2 &    0 &    0 &  142 &    2 &       $< 0.7$ &        $< 50$ &       $< 0.7$ \\
NGC 6656 (M22)  & $-8.5$ & $-1.70$ &   42 &    0 &    0 &    0 &   42 &    0 &       $< 2.4$ &       \nodata &       $< 2.4$ \\
NGC 6779 (M56)  & $-7.4$ & $-1.98$ &   49 &    1 &    0 &    0 &   47 &    1 &       $< 2.1$ &        $<100$ &       $< 2.1$ \\
NGC 6838 (M71)  & $-5.6$ & $-0.78$ &   32 &   12 &    0 &    0 &   25 &    9 &       $< 4.0$ &        $< 11$ &       $< 2.9$ \\
NGC 6864 (M75)  & $-8.6$ & $-1.29$ &  105 &   23 &    0 &    0 &   51 &   13 &       $< 2.0$ &       $< 7.7$ &       $< 1.6$ \\
NGC 7006        & $-7.7$ & $-1.52$ &   48 &    5 &    0 &    0 &   20 &    1 &       $< 5.0$ &        $<100$ &       $< 4.8$ \\
NGC 7078 (M15)  & $-9.2$ & $-2.37$ &  285 &   42 &    0 &    0 &  261 &   31 &       $< 0.4$ &       $< 3.2$ &       $< 0.3$ \\
NGC 7089 (M2)   & $-9.0$ & $-1.65$ &  358 &   11 &    0 &    0 &  317 &   11 &       $< 0.3$ &       $< 9.1$ &       $< 0.3$ \\
NGC 7099 (M30)  & $-7.4$ & $-2.27$ &  119 &    9 &    2 &    0 &  116 &    9 & $1.7 \pm 1.2$ &        $< 11$ & $1.6 \pm 1.1$ \\
Pal 13          & $-3.8$ & $-1.88$ &   10 &    0 &    0 &    0 &    5 &    0 &        $< 20$ &       \nodata &        $< 20$ \\
NGC 7492        & $-5.8$ & $-1.78$ &   15 &    4 &    0 &    0 &   10 &    0 &        $< 10$ &       \nodata &        $< 10$ \\
\tableline
Total           &        &         & 2057 &  170 &    4 &    2 & 1696 &  125 & $0.2 \pm 0.1$ & $1.6 \pm 1.1$ & $0.3 \pm 0.1$ \\
\enddata
\tablerefs{Cluster luminosities and metallicities are from the compilation of \protect \citet[][updated 2010]{har96} and references therein.}
\tablenotetext{a}{Stars observed with DEIMOS with Li detections or upper limits.}
\tablenotetext{b}{Includes detections of normal Li abundances and useful upper limits (where Li-richness would have been detected).}
\end{deluxetable*}

Stars with Li detections above the boundary are considered Li-rich.
Stars with Li detections or upper limits below the boundary are
Li-normal.  Upper limits above the boundary do not indicate whether
the star is Li-rich or Li-normal.  We calculated the frequency of
Li-rich stars as the number of Li-rich stars divided by the total
number of detections and ``useful'' upper limits.  If we were to raise
the boundary for Li-richness, fewer stars would be considered Li-rich,
and more upper limits would be considered useful, both of which would
decrease the Li-rich frequency.

Table~\ref{tab:gc} shows the Li-rich frequency for each GC in our
sample and for the combined sample of all \ncluster\ GCs.  The
``Members'' column shows stars that passed the membership criteria,
regardless of their Li abundances.  ``Li-rich'' shows stars with
$A({\rm Li})$ that exceed the boundary set by
Equation~\ref{eq:lirich}.  ``Li-normal'' includes stars with
detections or upper limits below the boundary.  The ``Li-Rich
Frequency'' is $(\text{Li-Rich})/(\text{Li-Rich} + \text{Li-Normal})$.
The error bars on the frequencies are Poissonian:
$\sqrt{(\text{Li-Rich})}/(\text{Li-Rich} + \text{Li-Normal})$.

Li-rich giants appear in M68, NGC~5053, M3, NGC~5897, and M30.  S232,
the more luminous giant in M68 was previously discovered by
\citet{ruc11}.  IV--101, the M3 giant discovered by \citet{kra99}, is
not in our sample.  These clusters do not appear remarkable in any way
other than hosting Li-rich giants.  Table~\ref{tab:gc} shows that
these clusters have typical luminosities and metallicities.

Two GCs host two Li-rich giants each.  M68 has one Li-rich RGB star
and one Li-rich AGB star, and M30 has two Li-rich RGB stars.  We
conducted $10^6$ random draws of stars from our sample in order to
test for the significance of this apparent clustering of Li-rich
giants.  We drew at least two RGB stars from the same GC in 32\% of
the trials, and we drew one RGB and one AGB star from the same GC in
46\% of the trials.  At least two stars of any type were drawn from
each of two or more GCs in 11\% of the trials.  Therefore, clustering
of Li-rich giants cannot be ruled out, but the significance is
marginal.

The fraction of Li-rich giants across all GCs in our sample is
$(\fractot \pm \fractoterr)\%$, notably less than the commonly quoted
1\%.  The statistics do not include IV--101 in M3 because it was not
included in our sample.  We obtained a longslit spectrum of this star
and confirmed its Li enhancement, but we did so because it was
pre-selected to be Li-rich.  In order to avoid biasing our results,
Table~\ref{tab:gc} includes only stars that were included in our
random sample.

\subsection{Stellar Evolutionary State}

\begin{figure*}[t!]
\centering
\includegraphics[width=0.85\textwidth]{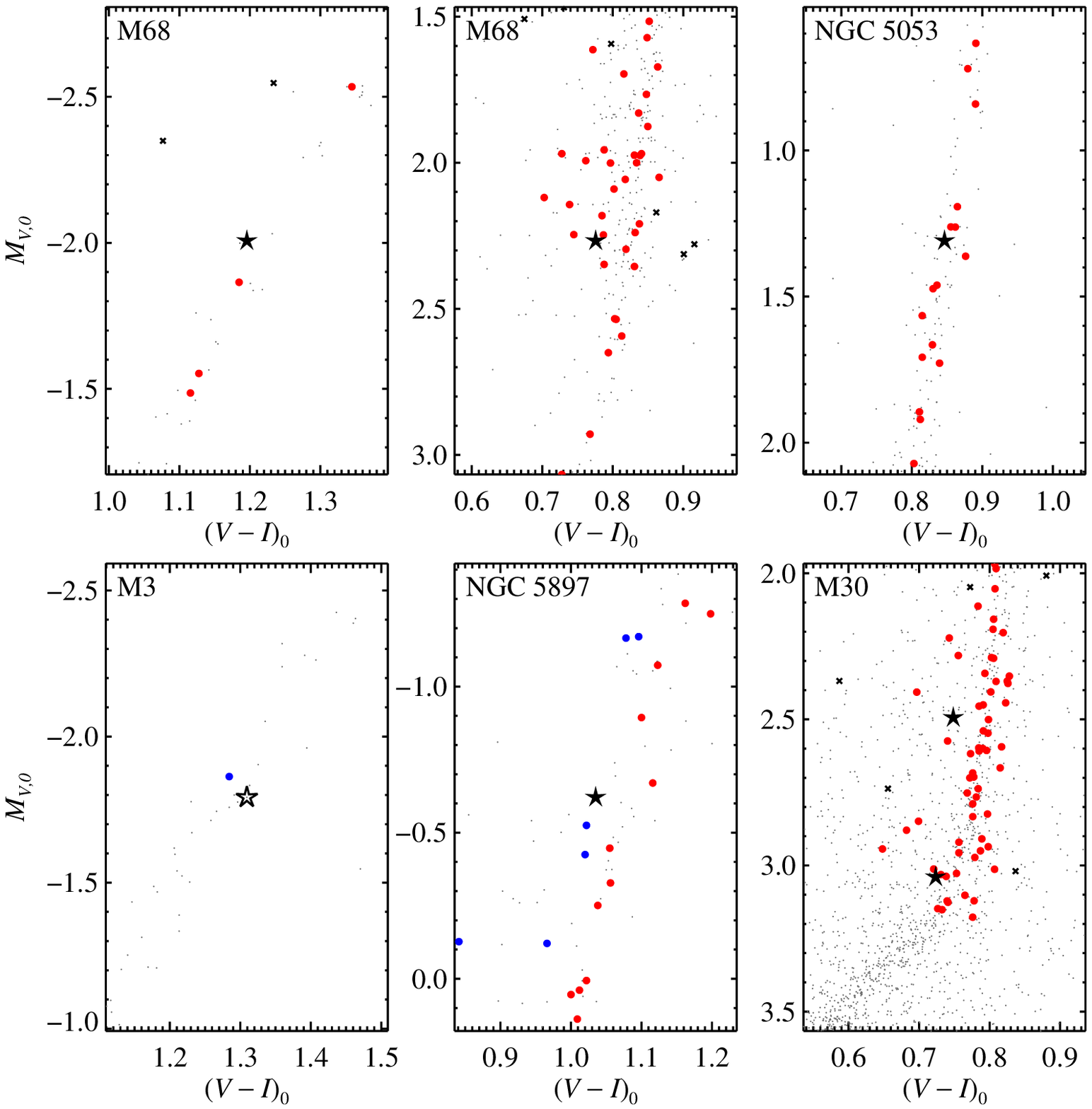}
\caption{Details of the CMDs for the four clusters with Li-rich
  giants.  Although M30 has two Li-rich giants, only one panel is
  shown because the two stars are close together on the CMD\@.
  Because M68 S232 (top left) and NGC~5897 Tes01--WF4--703 (bottom
  center) lie to the left of the RGB, we identified them as AGB stars.
  The hollow star indicates the Li-rich giant IV--101 in M3
  \citep{kra99}, which is not part of our sample.\label{fig:cmdli}}
\end{figure*}

It is useful to identify the stellar evolutionary phase of the Li-rich
giants in order to determine whether they could have generated the
extra Li themselves.  Figure~\ref{fig:cmdli} shows detail in the CMDs
around the Li-rich giants.  Four of the Li-rich giants are on the
lower RGB, but three giants are bright enough to be AGB stars: M68
S232, M3 IV--101 (not part of our sample), and NGC~5897 WF4--703.

The asymptotic nature of the AGB makes it difficult to assign RGB or
AGB status with complete confidence.  However, M68 S232 and NGC~5897
WF4--703 lie on the blue side of the giant branches.  M3 IV--101 lies
on the red side.  Therefore, the M68 and NGC~5897 stars are most
likely on the AGB, and the M3 star is most likely on the RGB\@.

Table~\ref{tab:gc} separates the Li-rich statistics into RGB and
AGB\@.  The Li-rich fraction is $(\fracrgb \pm \fracrgberr)\%$ for the
RGB (not including M3 IV--101) and $(\fracagb \pm \fracagberr)\%$ for
the AGB\@.  The statistics for RGB stars are more certain because we
observed over ten times more RGB stars than AGB stars.  Although the
sample sizes are small, we can estimate the probability that the
frequency of Li-rich RGB stars is the same as for Li-rich AGB stars.
The Poisson rate ratio test \citep{leh05} returns a $p$-value of 0.12
that the rate for Li-richness is the same for the RGB and AGB\@.
Hence, there is an 88\% chance that the two populations are different.
We consider this to be a marginally significant result.


\section{Discussion}
\label{sec:discussion}

In this section, we consider the three scenarios for Li enhancement
discussed in Section~\ref{sec:intro}: engulfment of a substellar
companion, self-generation, and mass transfer.  Although the
difference in Li-rich frequencies between the RGB and AGB is
marginally significant, different frequencies have significant
implications for the origin of Li enhancement.  We consider whether
each scenario would result in different frequencies for different
evolutionary states.

\subsection{Engulfment of a Substellar Companion}

The oldest known exoplanet resides in the GC M4 \citep{bac93,sig03}.
However, the exoplanet orbits a pulsar, and it is possible that the
pulsar captured the exoplanet from a main sequence star \citep{sig93}.
Other than this unusual scenario, exoplanets have limited
survivability in GCs due to dynamical interaction \citep{sig92}.
Indeed, searches for transiting exoplanets found none in the GCs
47~Tuc and $\omega$~Cen \citep{gil00,wel05,wel08}.

If the Li enrichment is due to rotationally induced mixing
\citep{den04} caused by an increase in angular momentum from the
engulfed companion, then the Li-rich stars should have higher rotation
rates.  Although there is some evidence for higher rotation rates
among metal-rich, Li-rich giants \citep{gui09}, there is no such
evidence for metal-poor field stars \citep{ruc11}.  Therefore,
metal-poor, Li-rich giants in \citeauthor{ruc11}'s sample seem not to
have generated their Li through rotationally induced mixing.

The low metallicities of GCs also disfavor exoplanet formation.  Gas
giant exoplanets are increasingly rare around more metal-poor stars
\citep{gon97,san04,fis05} with an occurrence rate of $<1\%$ for ${\rm
  [Fe/H]} < -1$ \citep{joh10}.  \citet{wan15} recently showed that gas
dwarfs and terrestrial exoplanets are more common around metal-rich
stars, although the dependence on metallicity is weaker than for gas
giants.  (We note that, in contrast to \citealt{wan15},
\citealt{buc12} and \citealt{nev13} found no correlation between host
metallicity and the occurrence of planets the size of Neptune or
smaller.)  All but one (M71) of the GCs in our sample have ${\rm
  [Fe/H]} < -1$.  Furthermore, the metallicities of the GCs known to
host Li-rich giants are ${\rm [Fe/H]} = -2.3$, $-2.3$, $-2.2$, $-1.9$,
and $-1.5$ (NGC~5053, M30, M68, NGC~5897, and M3).  The fact that the
more metal-rich GCs do not have higher occurrences of Li-rich giants
does not favor exoplanets---which tend to occur around metal-rich
stars---as the origin of the Li.

The engulfment of a hot Jupiter should occur on the lower RGB\@.  By
the time the star reaches the tip of the RGB, it will have attained
close to its maximum radius.  Any companion ingestion should happen
before then.  Thermohaline mixing above the luminosity function bump
will destroy any Li acquired from the exoplanet.  Therefore, ingestion
of a substellar companion cannot explain any Li-rich HB or AGB star.
The fact that we found a higher frequency of Li-richness on the AGB
than the RGB indicates that companion engulfment cannot be the
dominant cause of Li enhancement in giants.

\subsection{Self-Generation of Lithium}

All of the proposed methods for self-generation of Li invoke some
version of the Cameron--Fowler conveyor.  The conveyor can operate in
an intermediate-mass AGB star's thermal pulse because the convection
zone reaches the $pp$-II burning region.  However, activating the
conveyor in an RGB star requires non-canonical mixing.

We have already discussed rotation as one mechanism to induce mixing
\citep{den04}, but again, the rotation rates of metal-poor, Li-rich
field giants are not significantly higher than metal-poor, Li-normal
stars \citep{ruc11}.  The unremarkable rotation rates in
\citeauthor{ruc11}'s sample disfavor rotationally induced mixing not
only from companion engulfment but also from other sources, including
unusually high natal rotation rates.  It is possible that the Li-rich
giants in GCs have a different distribution of rotation rates than
Li-rich giants in the field, but the resolution of our spectra yields
line widths (45--50 km~s$^{-1}$) that are too large to detect rotation
in giants.  Regardless, \citet{pal06} were unable to produce Li-rich
giants in computational models of rotationally induced mixing.

Other scenarios typically pinpoint one evolutionary stage as the
impetus for non-canonical mixing.  For example, the mixing could occur
at the RGB luminosity function bump \citep{cha00,pal01} or at the He
core flash \citep{sil14,mon14}.  These scenarios predict that Li-rich
stars should appear only at or beyond these evolutionary stages.  Red
giants reach the bump before the core flash.  Therefore, giants that
have not reached the bump should not be Li-rich if the Li is to be
created at the bump or the He core flash.  However, we have found four
Li-rich red giants that are much less luminous than the RGB bump.  The
Li in these stars could not have been created from self-generation at
the bump or He core flash.

Even the Li-rich AGB stars in our sample pose problems for the
self-generation scenario.  Hot bottom burning is effective in creating
Li only in AGB stars more massive than 4~$M_{\sun}$ \citep{sac92}.
The predicted abundance of Li in the atmosphere of a 1~$M_{\sun}$ AGB
star with ${\rm [Fe/H]} = -2.3$ is only $A({\rm Li}) = 0.5$
\citep{kar10}.  However, extra mixing processes, possibly including
thermohaline mixing, can induce a Li overabundance in the atmospheres
of less massive AGB stars \citep{can10}.  The efficacy of this
mechanism at producing Li is still very uncertain.  The He core flash
is also a possible event for inciting extra deep mixing in these AGB
stars \citep{kum11,sil14,mon14}.

\subsection{Binary Mass Transfer}

The only well-understood sites for the generation of large
overabundances of Li are intermediate-mass (4--7~$M_{\sun}$) AGB stars
\citep{cam71,sac92,ven10}.  These stars did not live long enough to
survive in GCs until today.  However, an intermediate-mass AGB star
could donate its Li to a binary companion.  If the companion has a
mass less than 0.8~$M_{\sun}$, it would still be visible in the
cluster now.  The Li would have been transferred while the recipient
was still a dwarf star.  This is a favorable configuration for mass
transfer because the recipient has a much higher surface gravity than
the donor.  Although the first dredge-up would deplete some of the Li
as the recipient evolved onto the RGB, the dredge-up should deplete
the same fraction of surface Li in all stars.  Thus, a star enriched
in Li before the dredge-up would still appear enriched in Li after the
dredge-up when compared to stars at the same evolutionary stage.  The
abundance of Li in the atmosphere of an intermediate-mass AGB star can
reach up to $A({\rm Li}) = 4.5$ \citep{sac92}, and even higher Li
abundances have been observed \citep{del95}.  If the dwarf star
acquires that Li abundance early in its life, it could still have
$A({\rm Li}) \approx 3.3$---even more than the value we observed in
our four Li-rich RGB stars---after it passes through the first
dredge-up.

The binary frequency in GCs is around 5--10\%, but it could have been
a factor of two higher before dynamical interactions destroyed binary
systems \citep{ji13,ji15}.  Assuming that the binary mass ratio
distribution is flat \citep{bof10} and that the initial binary
frequency was 20\%, about 1.7\% of 4--7~$M_{\sun}$ stars would have
had companions of 0.8~$M_{\sun}$.  In a \citet{kro01} initial mass
function, 1.3\% of 0.08--100~$M_{\sun}$ stars lie in the mass range
4--7~$M_{\sun}$.  Therefore, about $1.7\% \times 1.3\% = 0.02\%$ of
0.8~$M_{\sun}$ stars once had a companion of 4--7~$M_{\sun}$.  This
frequency is an order of magnitude lower than the Li-rich RGB
frequency that we observed.  Thus, standard hot bottom burning
combined with mass transfer could explain about 10\% of Li-rich red
giants.

The majority of Li-rich giants need to be produced by a different
mechanism.  The recent discoveries of Li-rich, post-He flash stars
\citep{kum11,sil14,mon14} present another possibility for forming
Li-rich giants on the lower RGB\@.  These stars are not as Li-rich as
the predictions from hot bottom burning in intermediate-mass AGB
stars.  If these post-He flash stars were to be the source of Li for
stars on the lower RGB, they need to have donated the Li after the
recipient completed most of its first dredge-up.

Suppose that all low-mass, metal-poor giants experience one or more
short-lived phases of Li enhancement during or after the He core
flash.  Further suppose that all such giants with companions on the
lower RGB will transfer mass to that companion.  A 13~Gyr, $Z =
10^{-4}$ Yonsei-Yale isochrone \citep{dem04} predicts that a turn-off
star had an initial mass of 0.803~$M_{\sun}$ and a star at the RGB
bump had an initial mass of 0.812~$M_{\sun}$.  A star at the tip of
the RGB experiencing its first He core flash had an initial mass of
0.813~$M_{\sun}$.  If we again adopt a flat binary mass ratio
distribution but reduce the binary fraction to 10\% (the present
rather than the initial value), then about 0.1\% of stars at the He
core flash should have a binary companion on the lower RGB between the
turn-off and the bump.  This number is at the lower range of the
confidence interval that we measured.

The two main suppositions in this scenario deserve scrutiny.  First,
we supposed that all GC stars must produce Li on the HB or the AGB\@.
Although this seems like a severe requirement, it circumvents the
puzzle of why some giants appear Li-rich and others do not.  Instead,
all giants experience Li enhancement, but the enhancement is
short-lived.  This idea---that Li enhancement is rare because it is
short-lived---has been suggested several times before
\citep[e.g.,][]{del97,cha00}.  If we take our observation of the
frequency of Li-rich AGB stars, $\fracagb\%$, as representative of all
post-He flash stars, then this frequency is equal to the duty cycle of
Li enhancement.  The post-RGB (HB and AGB) lifetime of a 13~Gyr, $Z =
10^{-4}$ star is about 130~Myr \citep{mar08}.  In our scenario, all
such stars spend 2~Myr in a state of Li enhancement.

Second, we supposed that all post-RGB stars with companions on the
lower RGB will transfer mass.  The lower RGB stars have higher surface
gravities than the HB and AGB stars.  Therefore, if mass transfer
occurs, the HB or AGB star will be the donor.  Mass transfer becomes
less favorable as the mass difference between the two stars becomes
smaller because the secondary star will have a lower surface gravity.
After the first dredge-up and before the RGB bump, the mass of the
convective envelope in the recipient star is 0.23--0.40~$M_{\sun}$
\citep{kar03,kar14a}.  In order to enrich the star to $A({\rm Li}) =
2.5$, a typical value for the Li-rich RGB stars we observed, the star
would need to acquire $6 \times 10^{-10}~M_{\sun}$ of Li.  This could
be accomplished, for example, by accreting 0.03~$M_{\sun}$ of the
envelope of an HB or AGB star that enriched itself to $A({\rm Li}) =
3.5$.  This amount of mass transfer is not unreasonable.  In fact, a
single 0.8~$M_{\sun}$ AGB star with $Z = 10^{-4}$ can lose one quarter
of its mass in thermal pulses \citep{mar08}.

However, we also note that low-mass, low-metallicity AGB stars are
expected to generate $s$-process elements, like barium
\citep[e.g.,][]{kar14b}.  The \baiin\ line is apparent (but blended)
in the spectra of all of the Li-rich stars we observed.  The line does
not appear any stronger than it appears in the Li-normal stars of
similar stellar parameters.  Thus, it does not seem that either the
Li-rich AGB stars or the Li-rich RGB stars are very enhanced in the
$s$-process.  The lack of $s$-process enrichment in Li-rich GC giants
is consistent with metal-rich, Li-rich giants \citep{gar13}.  These
observations demand that the Li be created in the AGB star and/or
transferred to the RGB companion before thermal pulses can dredge up
large amounts of $s$-process material.

In our scenario, the donor star is no longer visible because it has
evolved past the AGB into a white dwarf, too faint to be observed.
However, the recipient RGB star is still in orbit around the white
dwarf.  We predict that Li-rich red giants should show radial velocity
variations.  Our prediction warrants multi-epoch spectroscopy for
those red giants that we found to be Li-rich.  We also note that HB or
AGB stars that generate Li themselves need not have a binary
companion.  In fact, only $\sim 10\%$ of those stars should have a
companion \citep{ji13,ji15}.  Therefore, it is not worrisome that
\citet{mon14} did not find radial velocity variations around the
Li-rich HB star they discovered in Trumpler~5.


\section{Summary}
\label{sec:summary}

Although giant stars should deplete Li, some Li-rich giants exist.
Some of these giants have Li abundances in excess of the primordial
value of the universe, which indicates that the Li is being created,
not merely saved from destruction.  We examined several proposals for
Li enrichment: engulfment of a substellar companion, self-generation,
and binary mass transfer.

A GC is the best site to study stellar evolution because it is a
single-age, nearly mono-metallic stellar population at a uniform
distance.  In a survey similar to that of \citet{pil00}, we searched
for Li-rich giants in GCs.  The primary difference between our studies
is that our sample size was seven times larger, thanks to the
multiplexing of Keck/DEIMOS\@.  We measured Li abundances or useful
upper limits in \nsig\ giants across \ncluster\ GCs.  We defined a
luminosity-dependent criterion for Li-richness, and we found that six
stars satisfied that criterion.  The overall frequency for Li-richness
in GCs is $(\fractot \pm \fractoterr)\%$.

Some of the proposed scenarios for Li enrichment predict different
frequencies of Li-rich giants at different evolutionary states.
Although many non-asteroseismological studies cannot confidently
disentangle the RGB and AGB for field stars, we exploited the
simplicity of GC stellar populations to assign RGB or AGB identities
to each star in our sample.  We found that the frequency of
Li-richness is $(\fracrgb \pm \fracrgberr)\%$ on the RGB and
$(\fracagb \pm \fracagberr)\%$ on the AGB\@.  A Poisson rate ratio
test returns an 88\% probability that these two frequencies differ.

We found no correlation between the Li-rich frequency and any property
of the GCs, including metallicity.  The fact that we found Li-rich
giants in extremely metal-poor GCs disfavors exoplanet engulfment as
the origin of the Li because exoplanets are not expected to form
around extremely metal-poor stars
\citep[e.g.,][]{gon97,san04,fis05,joh10,wan15}.  Furthermore, searches
for transiting exoplanets in GCs have found none
\citep{gil00,wel05,wel08}.

Our observations also disfavor self-generation of Li on the RGB\@.
All four of the Li-rich RGB stars we found are on the lower RGB, less
luminous than the RGB luminosity function bump.  Li-rich RGB stars in
GCs have also been found above the bump \citep{kra99,smi99}.  Most of
the proposed explanations for Li self-generation predict that it will
occur at the bump or after the He core flash
\citep[e.g.,][]{cha00,sil14}.  One exception is rotationally induced
mixing \citep{den04}, but one study with the spectral resolution to
measure rotation rates in metal-poor, Li-rich field giants
\citep{ruc11} found normal rotation rates.  Our discovery of four new
low-luminosity, Li-rich giants adds to the growing evidence that
Li-rich giants can be found at any evolutionary stage
\citep[e.g.,][]{mon11,mar13}.  The wide luminosity range of Li-rich
giants rules out most of the proposed scenarios for Li
self-generation.

\citet{kum11}, \citet{sil14}, and \citet{mon14} found Li-rich giants
on the HB, where stars have recently experienced a He core flash.
\citeauthor{sil14}\ supported the identification of their star as a He
core-burning, HB star with Kepler asteroseismology.  These discoveries
could suggest that the He core flash could incite extra deep mixing
that activates the Cameron--Fowler conveyor.  This explanation applies
only to stars on the HB or AGB, not the RGB\@.  However, a post-flash,
Li-rich star could transfer Li to a binary companion on the RGB\@.  If
all post-RGB stars experience brief phases of Li enhancement and all
such stars in binaries transfer mass to their companions, then we
expect 0.1\% of stars on the lower RGB to be Li-rich.  This frequency
is consistent with our observations, but it is at the low end of our
confidence interval.

Our proposed scenario for explaining Li-rich giants requires the
acceptance of several stringent assumptions, but it solves several
nagging problems.  First, low-luminosity Li-rich giants in GCs do not
need to be ``special'' compared to other Li-normal stars.  They merely
need to be in a mass transfer binary with a Li-enhanced star.  Second,
the rarity of the luminous Li-rich giants can be explained by a fast
duty cycle of Li enhancement.  These stars are ``special'' only
because we happened to observe them during a Li-rich phase.  Finally,
the Li enhancement does not require the accretion of a exoplanet,
whose existence is disfavored in GCs \citep{gil00,wel05,wel08,wan15}.

\acknowledgments This manuscript is dedicated to the memory of Bob
Kraft.  We are grateful to Bob for his mentorship and his
inspirational discovery of a Li-rich red giant in M3.

We thank the colloquium audience at Steward Observatory for helpful
feedback.  We also thank the anonymous referee for a detailed,
thoughtful report that improved this article.  A.J.Z., J.H., M.G., and
R.G.\ carried out their work through UCSC's Science Internship Program
for high school students.  P.G.\ acknowledges support from NSF grants
AST-1010039 and AST-1412648.

We are grateful to the many people who have worked to make the Keck
Telescope and its instruments a reality and to operate and maintain
the Keck Observatory.  The authors wish to extend special thanks to
those of Hawaiian ancestry on whose sacred mountain we are privileged
to be guests.  Without their generous hospitality, none of the
observations presented herein would have been possible.

{\it Facility:} \facility{Keck:II (DEIMOS)}

\clearpage
\begin{turnpage}
\renewcommand{\thetable}{\arabic{table}}
\setcounter{table}{3}
\begin{deluxetable}{llccccccccccc}
\tablewidth{0pt}
\tablecolumns{13}
\tablecaption{Stellar Properties and Lithium Abundances\label{tab:catalog}}
\tablehead{\colhead{GC} & \colhead{Name} & \colhead{RA (J2000)} & \colhead{Dec (J2000)} & \colhead{Branch} & \colhead{$V_0$} & \colhead{$(B-V)_0$} & \colhead{$(V-I)_0$} & \colhead{$M_V$} & \colhead{$T_{\rm eff}$ (K)} & \colhead{$\log g$ (cm~s$^{-2}$)} & \colhead{[Fe/H]} & \colhead{$A({\rm Li})$}}
\startdata
\cutinhead{Li-Rich}
M68      & Stet-M68-S232   & 12 39 33.44 & $-$26 43 13.3 & AGB & 13.07 &    1.06 &    1.20 &    $-2.01$ & 4462 & 0.98 & $-2.38 \pm 0.11$ & $  3.17 \pm 0.10$ \\
M68      & Stet-M68-S534   & 12 39 36.77 & $-$26 37 57.9 & RGB & 17.34 &    0.63 &    0.78 & \phs$2.27$ & 5488 & 3.15 & $-2.44 \pm 0.13$ & $  2.41 \pm 0.15$ \\
NGC 5053 & N5053-S79       & 13 16 38.75 & $+$17 41 48.2 & RGB & 17.51 &    0.56 &    0.85 & \phs$1.31$ & 5367 & 2.73 & $-2.30 \pm 0.11$ & $  2.72 \pm 0.14$ \\
NGC 5897 & Tes01-WF4-703   & 15 17 23.15 & $-$20 59 42.3 & AGB & 14.90 & \nodata &    1.04 &    $-0.62$ & 4774 & 1.70 & $-1.99 \pm 0.11$ & $  1.50 \pm 0.11$ \\
M30      & 132             & 21 40 09.50 & $-$23 09 46.4 & RGB & 17.60 & \nodata &    0.72 & \phs$3.04$ & 5640 & 3.54 & $-2.43 \pm 0.12$ & $  2.66 \pm 0.14$ \\
M30      & 7229            & 21 40 18.77 & $-$23 13 40.4 & RGB & 17.05 & \nodata &    0.75 & \phs$2.49$ & 5510 & 3.28 & $-2.32 \pm 0.11$ & $  2.87 \pm 0.13$ \\
\cutinhead{Li-Normal and Upper Limits}
NGC 288  & 206             & 00 52 15.51 & $-$26 41 04.7 & RGB & 15.13 & \nodata &    0.98 & \phs$0.37$ & 4744 & 2.16 & $-1.39 \pm 0.11$ & $< 0.42$          \\
NGC 288  & 1635            & 00 52 20.64 & $-$26 37 34.7 & RGB & 17.38 & \nodata &    0.82 & \phs$2.62$ & 5290 & 3.26 & $-1.38 \pm 0.11$ & $< 1.45$          \\
NGC 288  & 2133            & 00 52 27.93 & $-$26 37 08.6 & RGB & 17.16 & \nodata &    0.85 & \phs$2.40$ & 5228 & 3.14 & $-1.28 \pm 0.11$ & $< 1.55$          \\
NGC 288  & 2228            & 00 52 28.85 & $-$26 37 04.2 & RGB & 17.84 & \nodata &    0.82 & \phs$3.08$ & 5359 & 3.45 & $-1.33 \pm 0.11$ & $< 1.71$          \\
\nodata & \nodata & \nodata & \nodata & \nodata & \nodata & \nodata & \nodata & \nodata & \nodata & \nodata & \nodata & \nodata \\
\enddata
\tablecomments{The table lists Li-rich giants first.  The rest of the list is sorted by right ascension.  (This table is available in its entirety in a machine-readable form in the online journal.  A portion is shown here for guidance regarding its form and content.)}
\end{deluxetable}
\clearpage
\end{turnpage}

\clearpage
\LongTables
\renewcommand{\thetable}{\arabic{table}}
\setcounter{table}{1}
\begin{deluxetable*}{llcccccc}
\tablewidth{0pt}
\tablecolumns{8}
\tablecaption{DEIMOS Observations\label{tab:obs}}
\tablehead{\colhead{GC} & \colhead{Slitmask} & \colhead{Targets} & \colhead{UT Date} & \colhead{Airmass} & \colhead{Seeing} & \colhead{Exposures} & \colhead{Exp.\ Time} \\
\colhead{ } & \colhead{ } & \colhead{ } & \colhead{ } & \colhead{ } & \colhead{($''$)} & \colhead{ } & \colhead{(s)}}
\startdata
NGC 288         & n288     &       119 & 2008 Nov 24    & 1.9 &     0.9 & 3 &    1140 \\
                & 288l1    &       150 & 2014 Aug 27    & 1.5 &     0.6 & 5 &    5460 \\
                & 288l2    &       148 & 2014 Aug 28    & 1.5 &     1.0 & 4 &    4800 \\
                & 288l3    &       148 & 2014 Aug 29    & 1.5 &     0.6 & 4 &    4320 \\
                & 288l4    &       145 & 2014 Aug 30    & 1.5 &     0.8 & 3 &    4140 \\
                & 288l5    &       148 & 2014 Aug 31    & 1.5 &     0.8 & 3 &    4320 \\
Pal 2           & pal2     &    \phn45 & 2008 Aug 3\phn & 1.5 &     0.7 & 5 & \phn940 \\
NGC 1904 (M79)  & n1904\tablenotemark{a} &    \phn22 & 2006 Feb 2\phn & 1.4 & \nodata & 2 & \phn600 \\
                & ng1904   &    \phn40 & 2009 Feb 22    & 1.4 &     0.9 & 4 &    3600 \\
                & 1904l1   &    \phn98 & 2014 Aug 27    & 2.0 &     0.9 & 3 &    3420 \\
                & 1904l2   &    \phn97 & 2014 Aug 28    & 2.1 &     1.2 & 3 &    4200 \\
                & 1904l3   &    \phn96 & 2014 Aug 29    & 1.9 &     0.8 & 3 &    3360 \\
                & 1904l4   &    \phn96 & 2014 Aug 30    & 1.9 &     1.1 & 3 &    3840 \\
NGC 2419        & n2419\tablenotemark{a} &    \phn70 & 2006 Feb 2\phn & 1.2 & \nodata & 4 &    1200 \\
                & n2419c   &    \phn94 & 2009 Oct 13    & 1.2 &     0.6 & 2 &    2100 \\
                &          &           & 2009 Oct 14    & 1.2 &     0.5 & 3 &    2700 \\
                & n2419b   &       111 & 2012 Mar 19    & 1.1 &     0.7 & 3 &    2700 \\
NGC 4590 (M68)  & n4590a   &    \phn96 & 2011 Jun 2\phn & 1.5 &     0.7 & 3 &    2400 \\
                & n4590b   &    \phn96 & 2011 Jun 2\phn & 1.6 &     0.8 & 3 &    2400 \\
                & 4590l1   &    \phn95 & 2014 Jun 8\phn & 1.5 &     0.8 & 4 &    4800 \\
NGC 5024 (M53)  & ng5024   &    \phn40 & 2009 Feb 23    & 1.2 &     0.7 & 2 &    1600 \\
NGC 5053        & ng5053   &    \phn40 & 2009 Feb 23    & 1.5 &     0.9 & 3 &    3600 \\
NGC 5272 (M3)   & n5272c   &       132 & 2011 Jun 3\phn & 1.1 &     0.8 & 2 & \phn960 \\
NGC 5634        & n5634a   &    \phn62 & 2011 Jan 30    & 1.2 &     0.7 & 3 &    3700 \\
                & n5634b   &    \phn61 & 2011 Jun 2\phn & 1.1 &     0.7 & 3 &    3907 \\
NGC 5897        & 5897a    &       120 & 2011 Aug 6\phn & 1.4 &     0.8 & 3 &    1800 \\
                & 5897l1   &       117 & 2014 Jun 8\phn & 1.3 &     0.8 & 3 &    3600 \\
                & 5897l2   &       113 & 2014 Jun 8\phn & 1.4 &     0.8 & 3 &    3600 \\
                & 5897l3   &       111 & 2014 Jun 29    & 1.3 &     0.8 & 4 &    5400 \\
                & 5897l4   &       114 & 2014 Jun 30    & 1.4 &     0.7 & 5 &    6000 \\
NGC 5904 (M5)   & ng5904   &    \phn40 & 2009 Feb 22    & 1.1 &     0.6 & 4 &    3180 \\
Pal 14          & pal14a   &    \phn40 & 2011 Aug 6\phn & 1.3 &     1.2 & 3 &    3960 \\
NGC 6205 (M13)  & n6205    &    \phn93 & 2007 Oct 12    & 1.4 & \nodata & 3 & \phn900 \\
NGC 6229        & 6229a    &    \phn76 & 2011 Jun 3\phn & 1.2 &     0.7 & 3 &    3900 \\
NGC 6341 (M92)  & n6341a   &       149 & 2011 Jun 2\phn & 1.1 &     0.7 & 3 &    1800 \\
                & n6341b   &       150 & 2011 Jun 2\phn & 1.1 &     0.7 & 3 &    1800 \\
                & 6341l1   &       177 & 2014 Jun 8\phn & 1.1 &     0.8 & 3 &    3600 \\
                & 6341l2   &       174 & 2014 Jun 8\phn & 1.2 &     0.9 & 3 &    4200 \\
NGC 6656 (M22)  & n6656b   &    \phn64 & 2009 Oct 13    & 1.5 &     0.8 & 2 &    1800 \\
                &          &           & 2009 Oct 14    & 1.5 &     0.6 & 3 &    2220 \\
NGC 6779 (M56)  & 6779l1   &    \phn68 & 2014 Aug 27    & 1.1 &     0.7 & 4 &    4181 \\
                & 6779l2   &    \phn67 & 2014 Aug 27    & 1.0 &     0.8 & 4 &    4200 \\
                & 6779l3   &    \phn87 & 2014 Aug 27    & 1.1 &     0.7 & 4 &    4800 \\
NGC 6838 (M71)  & n6838    &       105 & 2007 Nov 13    & 1.1 &     0.6 & 3 & \phn900 \\
NGC 6864 (M75)  & 6864aB   &    \phn90 & 2011 Aug 5\phn & 1.4 &     1.1 & 4 &    4800 \\
                & 6864l1   &       120 & 2014 Jun 7\phn & 1.6 &     1.0 & 4 &    4800 \\
                & 6864l2   &       112 & 2014 Jun 7\phn & 1.4 &     0.9 & 4 &    4800 \\
NGC 7006        & n7006    &       105 & 2007 Nov 15    & 1.0 & \nodata & 2 & \phn600 \\
                & 7006a    &    \phn95 & 2011 Jun 3\phn & 1.1 &     0.7 & 3 &    5040 \\
NGC 7078 (M15)  & n7078    &    \phn64 & 2007 Nov 12    & 1.0 & \nodata & 1 & \phn300 \\
                &          &           & 2007 Nov 14    & 1.0 & \nodata & 2 & \phn600 \\
                & n7078d   &       164 & 2009 Oct 13    & 1.0 &     0.5 & 3 &    2700 \\
                & n7078e   &       167 & 2009 Oct 14    & 1.0 &     0.6 & 3 &    2700 \\
                & 7078l1B  &       175 & 2014 Aug 28    & 1.1 &     0.5 & 3 &    3600 \\
                & 7078l2B  &       171 & 2014 Aug 28    & 1.2 &     0.8 & 3 &    3600 \\
                & 7078l3B  &       166 & 2014 Aug 29    & 1.4 &     0.9 & 3 &    3600 \\
                & 7078l4B  &       167 & 2014 Aug 29    & 1.2 &     0.8 & 4 &    4320 \\
                & 7078l5B  &       169 & 2014 Aug 31    & 1.0 &     0.7 & 3 &    3600 \\
NGC 7089 (M2)   & n7089b   &    \phn91 & 2009 Oct 13    & 1.1 &     0.6 & 3 &    2700 \\
                & 7089c    &       142 & 2011 Jun 3\phn & 1.1 &     0.9 & 3 &    2340 \\
                & 7089l2   &       154 & 2014 May 28    & 1.2 & \nodata & 4 &    7200 \\
                & 7089l1   &       156 & 2014 May 29    & 1.2 & \nodata & 3 &    5400 \\
                & 7089m1   &       145 & 2014 Jun 8\phn & 1.1 &     0.8 & 4 &    5460 \\
                & 7089m2   &       147 & 2014 Jun 29    & 1.1 &     0.9 & 3 &    3060 \\
                & 7089l3   &       158 & 2014 Jun 30    & 1.1 &     1.0 & 3 &    2940 \\
                & 7089l4   &       155 & 2014 Aug 27    & 1.2 &     0.7 & 5 &    5820 \\
                & 7089l5   &       152 & 2014 Aug 30    & 1.4 &     0.9 & 3 &    3120 \\
                & 7089l6   &       149 & 2014 Aug 31    & 1.4 &     0.7 & 3 &    4320 \\
NGC 7099 (M30)  & n7099    &    \phn38 & 2008 Nov 26    & 1.4 &     0.7 & 4 &    1200 \\
                & 7099l1   &       165 & 2014 Aug 29    & 1.6 &     0.8 & 3 &    4320 \\
                & 7099l2   &       157 & 2014 Aug 29    & 1.4 &     0.8 & 3 &    4320 \\
                & 7099l3   &       153 & 2014 Aug 29    & 1.4 &     0.8 & 3 &    4320 \\
                & 7099l4   &       157 & 2014 Aug 30    & 1.4 &     0.8 & 3 &    3600 \\
                & 7099l5   &       156 & 2014 Aug 30    & 1.4 &     0.7 & 3 &    3600 \\
                & 7099l6   &       158 & 2014 Aug 30    & 1.5 &     0.7 & 3 &    3600 \\
                & 7099l7   &       158 & 2014 Aug 31    & 1.4 &     0.7 & 4 &    3900 \\
Pal 13          & pal13    &    \phn33 & 2009 Oct 13    & 1.5 &     0.6 & 2 &    1800 \\
                &          &           & 2009 Oct 14    & 1.5 &     0.7 & 2 &    1722 \\
NGC 7492        & n7492    &    \phn46 & 2007 Nov 15    & 1.3 & \nodata & 2 & \phn420 \\
\enddata
\tablenotetext{a}{Observations by \protect \citet{sim07}.}
\end{deluxetable*}

\end{document}